%% file: main.tex
\documentclass[a4paper]{scrartcl} 
\usepackage[T1]{fontenc}
\usepackage[utf8]{inputenc}
\usepackage{epsfig}
\usepackage{amssymb,amsmath,amsthm,mathrsfs,mathtools,amscd}
\usepackage{pgfplots}
\usepackage{xcolor}
\usepackage{float}
\usepackage{framed}
\usepgfplotslibrary{groupplots}
\pgfplotsset{compat=newest}
\usepackage[bookmarks=true]{hyperref}
\usepackage{listings}
\input{stylefile}
\input{smathcomabb}
\author{Maximilian Behr \and Peter Benner \and Jan Heiland}
\title{Example Setups of Navier-Stokes Equations with Control and Observation: Spatial Discretization and Representation via Linear-quadratic Matrix Coefficients}
\begin{document}
\maketitle
% \paragraph{Abstract} 
\begin{abstract}

We provide spatial discretizations of nonlinear incompressible Navier-Stokes equations with inputs and outputs in the form of matrices ready to use in any numerical linear algebra package.
We discuss the assembling of the system operators and the realization of boundary conditions and inputs and outputs. We describe the two benchmark problems \emph{driven cavity} and \emph{cylinder wake} and provide the corresponding data. The use of the data is illustrated by numerous example setups. The test cases are provided as plain \python~or \octmat~script files for immediate replication.
\end{abstract}
\tableofcontents
% \listoftodos
\input{intro}

\input{assembldata}

\input{bcs}

\input{dsystem}
\input{setups}
\input{controlsetups}
\input{numexa}

%%%%%%%%%%%%%%%%%%%%%%%%%%%%%%%%%%%%%%%%%%%%%%%%%%%%%%%%%%%%%%%%%%%%%%%%%%%%
% \renewcommand{\em}{\it} 

% \bibliographystyle{abbrv}
% \bibliography{bib_jh}
% \bibliography{nsequad}

\input{main.bbl}
% \input{bibas.bbl}
\end{document}

%%% Local Variables: 
%%% TeX-PDF-mode:t
%%% auto-fill-function:nil
%%% mode:auto-fill
%%% flyspell-mode:nil
%%% mode:flyspell
%%% ispell-local-dictionary: "american"
%%% End: 

%% file: stylefile.tex
\def\ifiss{\textsf{IFISS}}

\newcommand{\ddt}{\tfrac{\textup{d}}{\textup{d}t}}

\newcommand{\Nv}{\ensuremath{n}}
\newcommand{\Np}{\ensuremath{m}}
\newcommand{\Nu}{\ensuremath{{N_u}}}

\newcommand{\Ny}{\ensuremath{q}}

\newcommand{\fv}{\ensuremath{f}}
\newcommand{\bv}{\ensuremath{\mathbf v}}
\newcommand{\bw}{\ensuremath{\mathbf w}}
\newcommand{\bp}{\ensuremath{\mathbf p}}
\newcommand{\bu}{\bv}
\newcommand{\vu}{\bv}

\newcommand{\Cv}{{\ensuremath{\mathcal C_v}}}
\newcommand{\Cp}{{\ensuremath{\mathcal C_p}}}
\newcommand{\Cdv}{{\ensuremath{\mathcal C_v}}}
\newcommand{\Cdp}{{\ensuremath{\mathcal C_p}}}
\newcommand{\Bop}{{\ensuremath{\mathcal B}}}
\newcommand{\Bd}{\ensuremath{B}}
\newcommand{\Bbc}{{\ensuremath{B_{bc}}}}
\newcommand{\Abc}{{\ensuremath{A_{bc}}}}

\newcommand{\My}{\ensuremath{\mathcal M_{Y}}}

\providecommand{\inva}[1]{~\textup{d}#1}

\providecommand{\mlsti}[1]{\text{\lstinline$#1$}}
\providecommand{\conv}[2]{( #1 \cdot \nabla) #2}

\newcommand{\Ren}{\text{Re}}
\newcommand{\Rei}{\tfrac 1\Ren}

\newcommand{\cI}{\ensuremath{\mathcal I}}
\newcommand{\sI}{_{\text{I}}}
\newcommand{\sG}{_{\Gamma}}
\newcommand{\bui}{\bu\sI}
\newcommand{\bug}{\bu\sG}
\newcommand{\bbui}{\bar\bu\sI}
\newcommand{\bbug}{\bar\bu\sG}

\definecolor{color0}{rgb}{0.0860563360058, 0.238246924042, 0.305612363081}
\definecolor{color1}{rgb}{0.329277292634, 0.476284555658, 0.183715554976}
\definecolor{color2}{rgb}{0.81462453292, 0.495483165723, 0.575252593642}
\definecolor{color3}{rgb}{0.758718300801, 0.792206933547, 0.954386122191}

\newcommand\gaco{\ensuremath{\Gamma_{c_1}}}
\newcommand\gact{\ensuremath{\Gamma_{c_2}}}

\newcommand{\octmat}{\textsf{Octave/MATLAB}}
\newcommand{\python}{\textsf{Python}}

%% file: smathcomabb.tex
\DeclareMathOperator{\spann}{span}

\providecommand{\inva}[1]{\text{~d} #1}

\providecommand{\into}[0]{\int_\Omega}
\providecommand{\bbmat}{\begin{bmatrix}}
\providecommand{\ebmat}{\end{bmatrix}}

\providecommand{\andi}[0]{\quad \text{and} \quad}

%% file: intro.tex
\section{Introduction}
This work considers spatial discretizations of incompressible Navier-Stokes equations 
\begin{subequations}\label{eq:NSintro}
\begin{align}
	\frac{\partial v}{\partial t}+(v\cdot\nabla)v-\Rei\Delta v+\nabla p&=f\ \text{in}\ (0,t)\times\Omega, \label{eq:NSintro_mom}\\ 
		\nabla\cdot v&=0\ \text{in}\ (0,t)\times\Omega,
\end{align}
\end{subequations}
with controls and observations, which we assume to be parametrized by the \emph{Reynolds number} \Ren, posed on a bounded domain $\Omega \subseteq \mathbb R^{2}$ for time $t>0$, and equipped with an initial condition and boundary conditions. 

This manuscript is accompanied by data that represents spatial discretizations of \eqref{eq:NSintro} in various flow setups. The modelling of the flow problems, the discretizations, and the data representation comes with the following features and advantages
\begin{enumerate}
	\item The discretized nonlinearity is provided as an unfolded tensor $H$, that realizes $H(v,v)$ as $H\cdot v \otimes v$. On the one hand side, this leads to a pure matrix representation. On the other hand, the tensor representation is directly accessible to recent developments of model reduction for bilinear quadratic systems.
	\item The terms that are multiplied with $\Rei$  are kept separate, so that simulations with different \emph{Reynolds numbers} can be realized through a scaling of the corresponding matrices. 
	\item Nonhomogeneous boundary conditions are incorporated via a \emph{penalized Robin} approach. This allows us to model the control action via an linear input operator that appears in the right hand side of the momentum equation \eqref{eq:NSintro_mom} as in the case of distributed control; cf. \cite{BenH15a}.
\end{enumerate}

Altogether, these approaches provide a discrete representation of nonlinear Navier-Stokes equations with control and observations that can be readily used for simulations and system theoretical experiments not resorting to any finite element toolbox whatsoever.

The manuscript is structured as follows. First, we explain how the system coefficient matrices of the semi-discrete Navier-Stokes equations are assembled and explain the basic usage of the matrices in simulations. Then, we introduce the test setups for simulation or control of imcompressible flows. In the last part, we document the data files and the setup and results of some example configurations. 

\begin{figure}[h!]
  \begin{framed}
    \textbf{Code and Data Availability} \\
    The source code and the data of the implementations used to compute the presented results is available from:
    \begin{center}
			\href{https://doi.org/10.5281/zenodo.834940}{\texttt{doi:10.5281/zenodo.834940}}
    \end{center}
	under the \emph{MIT} license.
  % and is authored by: Maximilian Behr and Jan Heiland 
  \end{framed}
	\caption{Link to code and data.}\label{fig:linkcodndat}
\end{figure}

\section{Spatial Discretization of Navier-Stokes Equations}
We consider a general mixed finite element scheme to illustrate the derivation. For mathematical and practical details on finite element approximations of flows, see, e.g., \cite{ElmSW05, GreS00}. 

Let $V$ and $Q$ be the finite element space, namely finite-dimensional function spaces defined on a tesselation of $\Omega$, for the velocity and the pressure spanned by nodal bases like 
\begin{equation} \label{eq:vpfemspace}
	V = \spann \{\phi_i\}_{i=1}^{\Nv} \quad \text{and} \quad Q = \spann \{\psi_k\}_{k=1}^{\Np}, \quad \Nv, \Np \in \mathbb N.
\end{equation}

 Thus, we look for approximations $v$ and $p$ to the velocity and pressure solutions of \eqref{eq:NSintro} that are linear combinations of the basis functions, namely
\begin{equation} \label{eq:vpexpansion}
	v(t,x) = \sum_{j=1}^{\Nv} v_j(t) \phi_j(x) \quad \text{and} \quad p(t,x) = \sum_{k=1}^{\Np} p_k(t) \psi_k(x).
\end{equation}

In what follows, we will often omit the time dependency of the coefficients and the space dependency of the basis functions. Also, we will freely identify the discrete functions $v$ and $p$ with their coefficient vectors
\begin{equation*}
	\bv = 
	\begin{bmatrix}
		v_1 \\ v_2 \\ \vdots \\ v_{\Nv}
	\end{bmatrix}
	\quad \text{and} \quad
	\bp = 
	\begin{bmatrix}
		p_1 \\ p_2 \\ \vdots \\ p_{\Np}
	\end{bmatrix}
\end{equation*}
defined through the expansion in \eqref{eq:vpexpansion}.

%% file: assembldata.tex
\section{Assembling the Linear Operators}\label{sec:asslinop}

For the finite element spaces \eqref{eq:vpfemspace}, we assemble the finite dimensional approximations $A$ and $J$ to the diffusion and divergence operators $\Delta$ and $\nabla \cdot$ in the standard way:
\begin{equation} \label{eq:coeffs_diffnab}
	A = \left[\ \into \nabla \phi_i : \nabla \phi_j \inva x \right]_{\begin{subarray}{l}i=1,\ldots,\Nv \\ j=1,\ldots,\Nv\end{subarray}}\in \mathbb R^{\Nv\times\Nv} \quad \text{and}\quad J=\left[\ \into \psi_k\nabla\cdot\phi_j \inva x \right]_{\begin{subarray}{l}k=1,\ldots,\Np \\ j=1,\ldots,\Nv\end{subarray}}\in \mathbb R^{\Nv\times \Nv},
\end{equation}
where $:$ stands for the Frobenius inner product.

The mass matrix $M$ and the right hand side $f$ are assembled as follows:
\begin{equation}
	M = \left[\ \into \phi_i \cdot \phi_j \inva x \right]_{\begin{subarray}{l}i=1,\ldots,\Nv \\ j=1,\ldots,\Nv\end{subarray}}\in \mathbb R^{\Nv\times \Nv} \quad \text{and}\quad f_{\Nv}=\left[\ \into \phi_i\cdot \fv  \inva x \right]_{i=1,\ldots, \Nv} \in \mathbb R^{\Nv \times 1}.
	\label{eq:coeffs_massrhs}
\end{equation}

By now, we haven't taken into account the boundary conditions. This we will discuss in Section \ref{sec:bcs}.

\section{Assembling of the Trilinear Form}\label{sec:assnonlinop}
Consider two discrete velocities $v = \sum\limits_{j=1}^{\Nv} v_j \phi_j$ and $w= \sum\limits_{k=1}^{\Nv} w_k \phi_k$ with coefficient vectors $\bv$ and $\bw$ and the resulting convective term
\begin{equation*}
	\conv vw = \sum_{j=1}^{\Nv} \sum_{k=1}^{\Nv} v_jw_k \conv {\phi_j}{\phi_k}.
\end{equation*}
Testing this relation against the $i$-th basis function gives the relation
\begin{equation*}
	\sum_{j=1}^{\Nv} \sum_{k=1}^{\Nv} v_jw_k\into (\conv {\phi_j}{\phi_k}) \cdot \phi_i \inva x  = H_{i} (\bv \otimes \bw),
\end{equation*}
with the \emph{Kronecker product} $\otimes \colon \mathbb R^{\Nv} \times \mathbb R^{\Nv} \to \mathbb R^{\Nv^2}$ and where 
\begin{align}\label{eq:Hi}
	H_{i}:=  \bbmat \into \conv {\phi_j}{\phi_k}\cdot \phi_i \inva x \ebmat_{j=1,\ldots,\Nv;\ k=1,\ldots,\Nv} \in \mathbb R^{1\times \Nv^2}
\end{align}

Thus, the discrete approximation of the convection operator $v\mapsto \conv vv$ is given as 
\begin{equation} \label{eq:H}
	H\colon \bv \mapsto H (\bv \otimes \bv), \quad \text{where}\quad H=\bbmat H_1 \\ \vdots \\ H_\Nv \ebmat \in \mathbb R^{\Nv\times\Nv^2}
\end{equation}
and $H_i$ is as defined in \eqref{eq:Hi}.

%% file: bcs.tex
\section{Incorporation of Boundary Conditions}\label{sec:bcs}
We now explain how boundary conditions are treated in general and for the trilinear form in particular. 
\subsection{Dirichlet Boundary Conditions}
To begin with, we assume that we only have Dirichlet conditions for the velocity. This means that the value of $v$ at the boundary is known and, since we consider nodal bases, also the values of the coefficients of the discretized velocities associated with nodes at the boundary are known. In the theory of weak solutions, the test space is chosen to have a zero trace at the Dirichlet boundary. In practice, the basis functions $\phi_i$ that are associated with a node at a Dirichlet boundary, are left out when assembling the coefficient matrices like in \eqref{eq:coeffs_diffnab} and \eqref{eq:coeffs_massrhs}. Often it is more convenient to assemble the whole matrices and remove the lines corresponding to the boundary in a second step.

Let $\cI_\Gamma$ be the set of indices associated with the Dirichlet boundary nodes and let $\cI_I$ be the set that contains the indices of all other nodes, i.e. the nodes in the inner or at the boundary where Neumann conditions are posed. Assume that the entries of the solution vector are sorted such that 
\begin{equation}\label{eq:innerbcnodes}
	\bu = \bar \bu\sI + \bar \bu\sG :=
	\begin{bmatrix}
		\bu\sI \\ 0
	\end{bmatrix} +
	\begin{bmatrix}
		0 \\ \bu_\Gamma
	\end{bmatrix} ,
\end{equation}
where $\bu\sI$ and $\bu\sG$ are the vectors of coefficients associated with $\cI\sI$ and $\cI\sG$. 

Let $[\cdot]\sI$ denote the operation of restricting a matrix or vector to the inner nodes with respect to degrees of freedom (columns) and the posed conditions (rows). For example, assume that $A$ is assembled without considering boundary conditions. Then, we resolve given Dirichlet boundary conditions via the relation
\begin{equation}\label{eq:resobcnodes}
	[A\bu]\sI = [A]\sI\bui + [A\bbug]\sI. \\
\end{equation}
We will use the abbreviation $A\sI:= [A]\sI$.

For the quadratic term, we first compute
\begin{align*}
	H(\bu\otimes \bu) 	&= 
	H(\bbui \otimes \bbui) + H(\bbui\otimes \bbug) + H(\bbug\otimes \bbui) + H(\bbug\otimes \bbug). 		
\end{align*}
Because of the zero entries in $\bbui$ and $\bbug$, it holds that 
\begin{equation}\label{eq:Hwbc}
	[H(\bu\otimes \bu)]\sI \phantom{:}= H\sI(\bui \otimes \bui) + [L_1(\bbug)]\sI\bui+ [L_2(\bbug)]\sI\bui + [H(\bbug\otimes\bbug)]\sI,
\end{equation}
		where
		\begin{subequations}\label{eq:Hwbc_linparts}
	\begin{align}
		L_1(\bbug) & = 
		\left[ \sum\limits_{k \in \cI\sG} [\bug]_k  \into (\conv {\phi_j}{\phi_k})\cdot \phi_i\inva x \right]_{\begin{subarray}{l}i=1,\ldots,\Nv \\ j=1,\ldots,\Nv\end{subarray}},\\
			L_2(\bbug) & = 
			\left[\sum\limits_{k \in \cI\sG} [\bug]_k \into (\conv {\phi_k}{\phi_j})\cdot \phi_i \inva x \right]_{\begin{subarray}{l}i=1,\ldots,\Nv \\ j=1,\ldots,\Nv\end{subarray}}.
			\end{align}
\end{subequations}

Thus, the consideration of Dirichlet boundary conditions in the quadratic formulation can be done as follows
\begin{enumerate}
	\item Assemble $H=H\sI$ like in \eqref{eq:H} for the inner nodes both in test and trial space.
	\item Assemble the linearized convection matrix $L(\bbug)$ with $\bbug$ as convection velocity like $L_1(\bbug)$ and $L_2(\bbug)$ in \eqref{eq:Hwbc_linparts} and restrict it to the inner nodes.
	\item Assemble the contribution $H(\bbug \otimes\bbug)$ to the source term via

		\begin{equation*}
		H(\bbug\otimes \bbug) = 
		\left[ \sum\limits_{j \in \cI\sG}\sum\limits_{k \in \cI\sG} [\bug]_j[\bug]_k \into (\conv {\phi_j}{\phi_k})\cdot \phi_i\inva x \right]_{\begin{subarray}{l}i=1,\ldots,\Nv\end{subarray}}
		\end{equation*}
		and restrict it to the inner nodes.
\end{enumerate}
As the result, one has the coefficient for $\bu\sI\otimes \bu\sI$, an additional coefficient matrix for $\bu\sI$, and a correction for the source term.

\subsection{Natural Boundary Conditions}
Boundary conditions that contain the normal derivative of $v$ are often called \emph{natural boundary condtions} since they like naturally are resolved in the finite element discretization. In view of assembling the coefficients, nodes associated with natural boundary are simply treated like inner nodes. However, one has to consider additional source terms, that derive from the partial integration that led to the forms in \eqref{eq:coeffs_diffnab}. 

Let $\Gamma_N$ denote the part of the boundary where boundary conditions other than of Dirichlet type are applied. Then the additional source term is assembled as
\begin{equation}
	g_\Nv 	= \left[ \int_{\Gamma_N} (p \vec n - \nu \frac{\partial v}{\partial \vec n}) \cdot \phi_i \inva x \right]_{i=1,\dotsc,\Nv}
			= \left[ \int_{\Gamma_N} g \cdot \phi_i \inva x \right]_{i=1,\dotsc,\Nv},
	\label{eq:normalbcsterm}
\end{equation}
where $\vec n$ is the outward normal vector. We will frequently employ \emph{do-nothing} conditions at the outflow boundary, i.e. we set
\begin{equation} \label{eq:donothingcds}
p \vec n - \nu \frac{\partial v}{\partial \vec n} = 0, \quad \text{on }\Gamma_N,
\end{equation}
so that $g_\Nv$ is just zero.

%% file: dsystem.tex
\section{The Spatially Discretized System}
We consider the Navier-Stokes equations
\begin{subequations}\label{eq:NSsys}
\begin{align}
  \frac{\partial \vu}{\partial
    t}-\Rei\Delta\vu+(\vu\cdot\nabla)\vu+\nabla \bp&=f_\Nv, \label{eq:NSmom} \\
		\nabla\cdot\vu&=0, \label{eq:NScont}
		\intertext{on the domain $\Omega$ with Dirichlet or \emph{do-nothing} boundary conditions that we write as }
\gamma \vu &= \vu_\Gamma	.  \label{eq:NSbc}
\end{align}
\end{subequations}

With the splitting of the velocity ansatz space into inner and boundary nodes as defined in \eqref{eq:innerbcnodes}, with the convention \eqref{eq:resobcnodes}, and with the formula for the convective term \eqref{eq:Hwbc}, a standard finite element discretization of \eqref{eq:NSsys} for the velocity at the inner nodes $\bui$ and the discrete pressure $\bp$ can be written as
\begin{subequations} \label{eq:NS_bili_disc}
\begin{align} 
	M\sI \ddt \bui + H\sI (\bui \otimes \bui) + [L_1(\bbug)+L_2(\bbug)+ \Rei A]\sI \bui - J\sI^T\bp &= [f_\Nv]{\sI}-[g_\Nv]{\sI} - \Rei [A\bbug]\sI- [H(\bbug\otimes\bbug)]\sI, \\
	J\sI\bui &= -[J\bbug]\sI.
\end{align}
\end{subequations}
For a given setup, $\bbug$ is defined through \eqref{eq:NSbc}, and, thus, we can define system \eqref{eq:NS_bili_disc} completely by means of the matrices and vectors
\begin{table}[h]
	\centering
	\begin{tabular}{lcl}
		\lstinline$M$ & $\leftrightarrow$ & $M\sI$\\
		\lstinline$A$ & $\leftrightarrow$ & $A\sI$\\
		\lstinline$H$ & $\leftrightarrow$ & $H\sI$\\
		\lstinline$L1$& $\leftrightarrow$ &$[L_1(\bbug)]\sI $\\
		\lstinline$L2$& $\leftrightarrow$ &$[L_2(\bbug)]\sI $\\
		\lstinline$J$ & $\leftrightarrow$ & $J\sI$\\
		\lstinline$J.T$ & $\leftrightarrow$ & $J\sI^T$
	\end{tabular}
	\quad and \quad
	\begin{tabular}{lcl}
		\lstinline$fv$ & $\leftrightarrow$ & $[f_\Nv]\sI$\\
		\lstinline$gv$ & $\leftrightarrow$ & $[g_\Nv]\sI$\\
		\lstinline$fv_diff$ & $\leftrightarrow$ & $[A\bbug]\sI$\\
		\lstinline$fv_conv$ & $\leftrightarrow$ & $[N(\bbug)\bbug]\sI$\\
		\lstinline$fp_div$ & $\leftrightarrow$ & $[J(\bbug)]\sI$
	\end{tabular}
	\caption{Parameters defining the spatially discretized Navier-Stokes equations \eqref{eq:NS_bili_disc}.}
	\label{tab:params}
\end{table}

up to a scaling of \lstinline$A$ and \lstinline$fv_diff$ by $\Rei$.

\section{Simulating the System}
Once one has the coefficients listed in Table \ref{tab:params} at hand, the implementation of a simulation can be done in any numerical linear algebra package. To illustrate the usage, we address a few important points of a simulation.
\subsection{Steady-state Solutions}
For a \emph{Reynolds number} \lstinline$RE$, the steady-state Stokes solution \lstinline$[v, p]$ is defined through
\begin{equation}\label{eq:ssstokes}
	\begin{bmatrix}
		\mlsti{1./RE*A} & \mlsti{-J.T} \\
		\mlsti J & \mlsti 0
	\end{bmatrix} 
	\mlsti *
	\begin{bmatrix}
		\mlsti v \\ \mlsti p
	\end{bmatrix}
	\mlsti{ != }
	\begin{bmatrix}
		\mlsti{fv - 1./RE*fv_diff} \\
		\mlsti{-fp_div}
	\end{bmatrix}
\end{equation}
and the steady-state Navier-Stokes solution is given as the solution of
\begin{equation*}
	\begin{bmatrix}
		\mlsti{1./RE*A+L1+L2} & \mlsti{-J.T} \\
		\mlsti J & \mlsti 0
	\end{bmatrix} 
	\mlsti *
	\begin{bmatrix}
		\mlsti v \\ \mlsti p
	\end{bmatrix}
	\mlsti{ != }
	\begin{bmatrix}
		\mlsti{fv - H*kron(v, v)- 1./RE*fv_diff - fv_conv} \\
		\mlsti{-fp_div}
	\end{bmatrix},
\end{equation*}
where \lstinline$!=$ means that a solver has to be applied and where \lstinline$kron$ is the function that computes the Kronecker product. Note that \lstinline$v$ represents the velocity solution only at the inner nodes, while the values of the boundary nodes are defined by the boundary conditions. 

\subsection{Linearizations}
We consider linearizations as they occur in the modelling of optimal control systems or during the numerical solution of the nonlinear state equations. 

In the nonlinear setting the convection is modelled through
\begin{lstlisting}
 L1*v + L2*v + H*kron(v, v) + fv_conv
\end{lstlisting}

If $a$ is a linearization point that in particular fulfills the boundary conditions and if \lstinline$a$ is its approximation, 
then the \emph{Oseen} or \emph{Picard} linearization $(u\cdot \nabla) u \leftarrow (a\cdot \nabla)u$ is realized as 
\begin{lstlisting}
 H*kron(a, v) + L1*v + L2*a + fv_conv
\end{lstlisting}
and the \emph{Newton} linearization $(u\cdot \nabla) u \leftarrow (a\cdot \nabla)u+(u\cdot \nabla)a-(a\cdot \nabla)a$ as
\begin{lstlisting}
 H*kron(a, v) + H*kron(v, a) - H*kron(a, a) + L1*v + L2*v + fv_conv
\end{lstlisting}

In the frequent case, that one decomposes \lstinline$v != vs + vd$, where \lstinline$vs$ is the steady-state Navier-Stokes solution, the linearized system for the difference state \lstinline$vd$ to be solved reads
\begin{lstlisting}
 M*vd.dt + A*vd + H*kron(vs, vd) + H*kron(vd, vs) - J.T*p != 0
 # and, simultaneously,
 J*vd != 0
\end{lstlisting}
where \lstinline$vd.dt$ denotes the time derivative. Note that, because the difference state is zero at the boundary, in particular the terms \lstinline$L1+L2$ and \lstinline$f_conv$ do not appear.

The matrices \lstinline$H1$ and \lstinline$H2$ representing the linear functions \lstinline$H*kron(a, v) =: H1*v$ and \lstinline$H*kron(v, a) =: H2*v$ can be computed columnwise:
\begin{lstlisting}
	H1 = [H*kron(e_i, a) for i in range(Nv)]
	H2 = [H*kron(a, e_i) for i in range(Nv)]
\end{lstlisting}
where \lstinline$a[i]$ is the $i$-th component of \lstinline$a$ and \lstinline$e_i$ is the $i$-th unit basis vector.

\subsection{Time Integration}\label{sec:timint}
Assume that the current velocity approximation \lstinline$v_old$ is given. Then one can apply the implicit-explicit (IMEX) Euler scheme to advance the velocity by a time step of size \lstinline$dt$ by solving

\begin{align*}
	& \begin{bmatrix}
		\mlsti{M + dt*(A+L1+L2)} & \mlsti{-dt*J.T} \\
		\mlsti J & \mlsti 0
	\end{bmatrix} 
	\mlsti *
	\begin{bmatrix}
		\mlsti{v_new} \\ \mlsti p
	\end{bmatrix}
	\mlsti{ != } \\
	&\quad\quad\quad\quad \begin{bmatrix}
		\mlsti{M*v_old + dt*(fv - H*kron(v_old, v_old)- fv_diff - fv_conv)} \\
		\mlsti{-fp_div}
	\end{bmatrix}.
\end{align*}

%% file: setups.tex
\section{Setups}
We describe the setups for which the coefficient matrices can be downloaded.

For all considered setups we describe the geometrical properties, the boundary conditions. The provided matrices are obtained by finite element discretizations as provided by \emph{FEniCS} \cite{LogOeRW12}. Depending on the geometry, we use uniform and nonuniform grids of different coarseness. As finite element pair we choose \emph{Taylor-Hood} $P_2-P_1$ finite elements \cite{TayH73}.

% or \emph{Crouzeix-Raviart} \cite{CroR73} elements which are popular and stable and that we denote as \TH~and \CR.

\subsection{Lid Driven Cavity}\label{sec:drivcavplain}

The two-dimensional driven cavity is a well investigated and common benchmark problem \cite{ShaD00}. As the model, one considers the incompressible Navier-Stokes equations on a square with homogeneous Dirichlet boundary conditions for the velocity except from the upper boundary $\Gamma_0$ -- the lid -- where the tangential velocity component is set to $1$. In the setup of the regularized driven cavity the tangential velocity set to smoothly approach zero towards the edges in order to comply with the zero conditions on the other walls. The regularized and standard driven cavity is a test problem of the \octmat~FE package \ifiss~\cite{ifiss}.

We consider the non regularized driven cavity on the unit square $\Omega = (0, 1)\times(0, 1)$ and its boundary $\Gamma$ with boundary conditions
\begin{equation*}
	\gamma v :
	\begin{cases}
		v = [1, 0 ]^T, \quad \text{on }\Gamma_0, \\
		v = [0, 0 ]^T, \quad \text{elsewhere},
	\end{cases}
\end{equation*}
cf. Fig. \ref{fig:drivcav}. For the spatial discretization, we use a uniform grid as illustrated in Fig. \ref{fig:drivcav} with a parameter \lstinline$N$ that describes the number of segments that one boundary is divided into.

\begin{figure}[tb]
	\newlength\figureheight
	\newlength\figurewidth
	\setlength\figureheight{6cm}
	\setlength\figurewidth{6cm}
	\input{pics/drivcav.tikz}
	\input{pics/drivcav_grid.tikz}
	\caption{Illustration of the velocity magnitude of the driven cavity problem and the domain of control and observation $\Omega_c=[0.4,0.6]\times [ 0.2,0.3]$ and $\Omega_o = [0.45,0.55] \times [0.5,0.7]$. The second plot shows a triangulation of the domain for \lstinline$N=10$.}
	\label{fig:drivcav}
\end{figure}
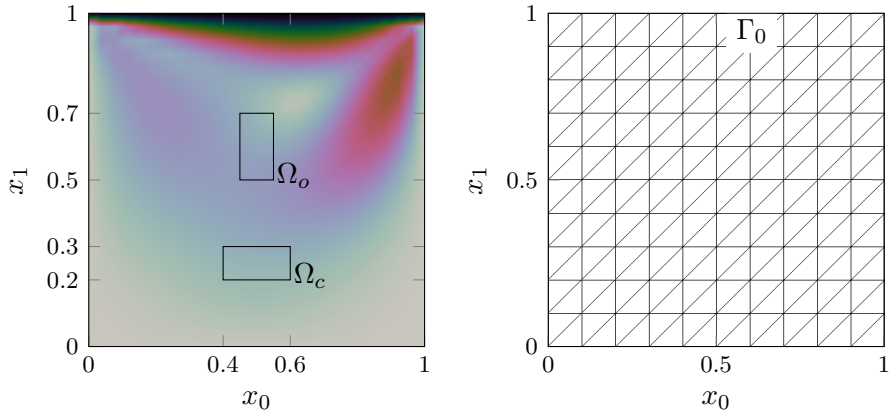

% As FE pairs we employ \TH~and \CR. 
Note that for flow setups with only Dirichlet boundary conditions for the velocity, like the driven cavity, the pressure is defined up to a constant. In the numerical approximation this leads to a rank deficiency of the divergence and the gradient matrices \lstinline$J$ and \lstinline$JT$. For iterative schemes this does not pose a problem.
However, in order to have the coefficient matrix in \eqref{eq:ssstokes} invertible, one can \emph{pin} the pressure to zero at one node of the discretization by simply removing the corresponding row and column of \lstinline$J$ and \lstinline$JT$.

\subsection{Cylinder Wake}\label{sec:cylwakeplain}

We consider the two dimensional wake of a cylinder which is a popular flow benchmark problem, a subject for simulations, and a test field for flow control \cite{Wil96, NoaAMTT03, BenH15}. 

The computational domain is as depicted in Fig. \ref{fig:cylwake}. 

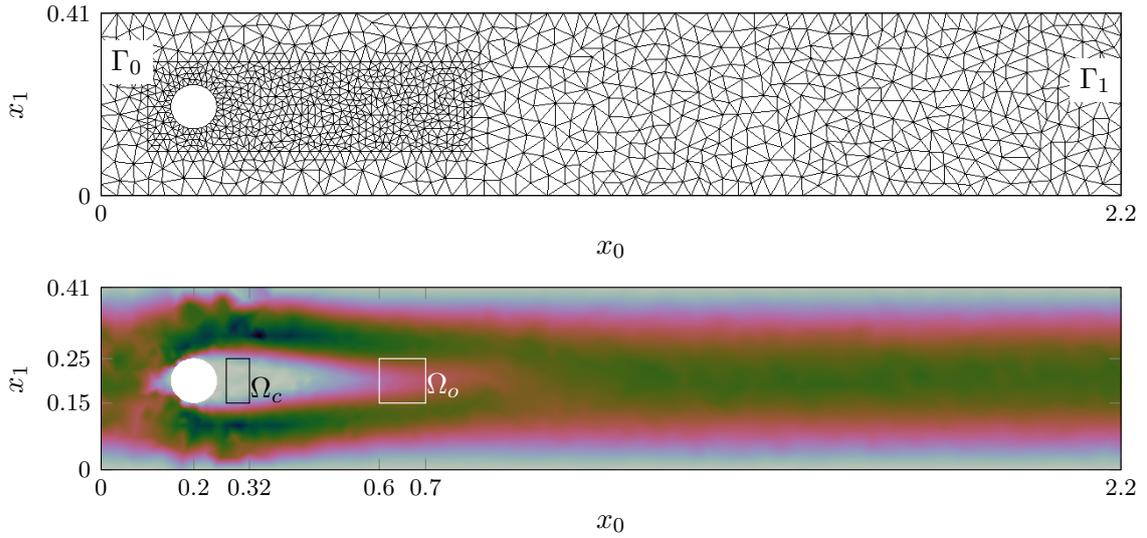
\begin{figure}
	\setlength\figureheight{4cm}
	\setlength\figurewidth{15cm}
	\input{pics/cylwake_grid.tikz}
	\input{pics/cylwake.tikz}
	\caption{Illustration of geometrical setup including the domains of distributed control and observation and of the velocity magnitude for the cylinder wake.} 
	%with \lstinline$nu=2E-3$, with \CR~on a spatial grid with parameter \lstinline$N=2$ (as illustrated above), started at the steady-state Stokes solution, advanced with a time step length \lstinline$dt=1./256$ to time \lstinline$te=1.$ using the semi-implicit Euler scheme as described in Section \ref{sec:timint}.}
	\label{fig:cylwake}
\end{figure}

We consider the incompressible Navier-Stokes equations on the domain as illustrated in Fig. \ref{fig:cylwake} with boundary $\Gamma$ with boundary conditions as follows, at the inflow $\Gamma_0$ we prescribe a parabolic velocity profile through the function 
$$g(s) = 4\bigl( 1 - \frac{s}{0.41}\bigr)\frac{s}{0.41} .$$
We impose \emph{do-nothing} conditions, cf. \eqref{eq:donothingcds}, at the outflow $\Gamma_1$ and \emph{no-slip}, i.e. zero Dirichlet conditions, at the upper and the lower wall of the channel and at the cylinder periphery. 
\begin{equation*}
	\gamma (v, p) \colon
	\begin{cases}
		v = [g(x_1), 0 ]^T, \quad \text{on }\Gamma_0, \\
	p \vec n - \Rei \frac{\partial v}{\partial \vec n} = [0, 0]^T, \quad \text{on }\Gamma_1, \\
		v = [0, 0 ]^T, \quad \text{elsewhere on the boundary}.
	\end{cases}
\end{equation*}

% Again, we use \TH~and \CR~elements to assemble a spatially discretized system on nonuniform grids of various coarseness described by the parameter \lstinline$N$.

%% file: pics/drivcav.tikz
% This file was created by matplotlib v0.1.0.
% Copyright (c) 2010--2014, Nico Schlömer <nico.schloemer@gmail.com>
% All rights reserved.
% 
% The lastest updates can be retrieved from
% 
% https://github.com/nschloe/matplotlib2tikz
% 
% where you can also submit bug reports and leavecomments.
% 
\begin{tikzpicture}

\begin{axis}[
xmin=0, xmax=1,
ymin=0, ymax=1,
width=\figureheight,
height=\figurewidth,
axis on top,
xtick={0,0.4,0.6,1},
ytick={0,.2,.3,.5,.7,1},
tick label style={font=\footnotesize},
xlabel={$x_0$},
ylabel={$x_1$},
xlabel near ticks,
ylabel near ticks
]
\addplot graphics [includegraphics cmd=\pgfimage,xmin=0, xmax=1, ymin=0, ymax=1] {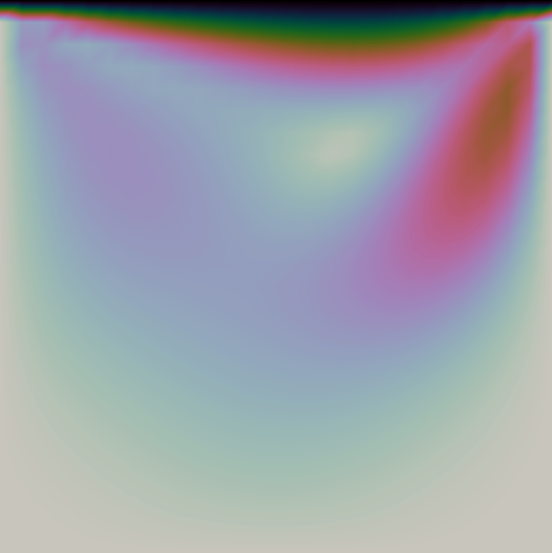};
\path [draw=black, fill opacity=0] (axis cs:13,1)--(axis cs:13,1);

\path [draw=black, fill opacity=0] (axis cs:0.05,13)--(axis cs:0.05,13);

\path [draw=black, fill opacity=0] (axis cs:13,1.38777878078145e-17)--(axis cs:13,1.38777878078145e-17);

\path [draw=black, fill opacity=0] (axis cs:0,13)--(axis cs:0,13);

\draw [color=black] (axis cs:0.4,0.2) rectangle (axis cs:0.6,0.3);
\node at (axis cs:0.58,0.28) [color=black, anchor=north west] {$\Omega_c$};
\draw [color=black] (axis cs:0.45,0.5) rectangle (axis cs:0.55,0.7);
\node at (axis cs:0.53,0.58) [color=black, anchor=north west] {$\Omega_o$};
\end{axis}

\end{tikzpicture}

%% file: pics/drivcav_grid.tikz
% This file was created by matplotlib v0.1.0.
% Copyright (c) 2010--2014, Nico Schlömer <nico.schloemer@gmail.com>
% All rights reserved.
% 
% The lastest updates can be retrieved from
% 
% https://github.com/nschloe/matplotlib2tikz
% 
% where you can also submit bug reports and leavecomments.
% 
\begin{tikzpicture}

\begin{axis}[
xmin=0, xmax=1,
ymin=0, ymax=1,
width=\figureheight,
height=\figurewidth,
axis on top,
xtick={0,.5,1},
ytick={0,.5,1},
tick label style={font=\footnotesize},
xlabel={$x_0$},
ylabel={$x_1$},
xlabel near ticks,
ylabel near ticks
]
\addplot graphics [includegraphics cmd=\pgfimage,xmin=0, xmax=1, ymin=0, ymax=1] {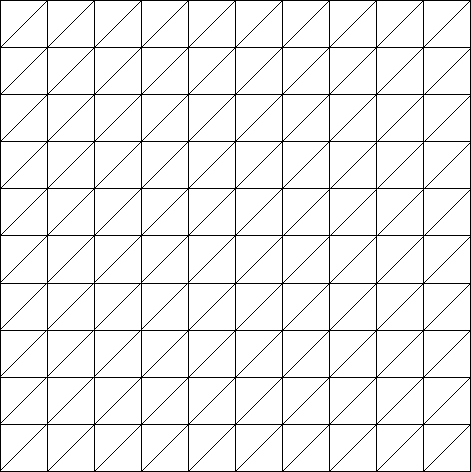};

\path [draw=black, fill opacity=0] (axis cs:13,1)--(axis cs:13,1);

\path [draw=black, fill opacity=0] (axis cs:0.05,13)--(axis cs:0.05,13);

\path [draw=black, fill opacity=0] (axis cs:13,1.38777878078145e-17)--(axis cs:13,1.38777878078145e-17);

\path [draw=black, fill opacity=0] (axis cs:0,13)--(axis cs:0,13);

\node at (axis cs:0.53,0.88) [color=black, fill=white, anchor=south west] {$\Gamma_0$};

\end{axis}

\end{tikzpicture}

%% file: pics/cylwake_grid.tikz
% This file was created by matplotlib v0.1.0.
% Copyright (c) 2010--2014, Nico Schlömer <nico.schloemer@gmail.com>
% All rights reserved.
% 
% The lastest updates can be retrieved from
% 
% https://github.com/nschloe/matplotlib2tikz
% 
% where you can also submit bug reports and leavecomments.
% 
\begin{tikzpicture}

\begin{axis}[
xmin=0, xmax=2.2,
ymin=0, ymax=.41,
width=\figurewidth,
height=\figureheight,
axis on top,
xtick={0,2.2},
ytick={0,.41},
tick label style={font=\footnotesize},
xlabel={$x_0$},
ylabel={$x_1$},
xlabel near ticks,
ylabel near ticks
]
\addplot graphics [includegraphics cmd=\pgfimage,xmin=0, xmax=2.2, ymin=0, ymax=0.41] {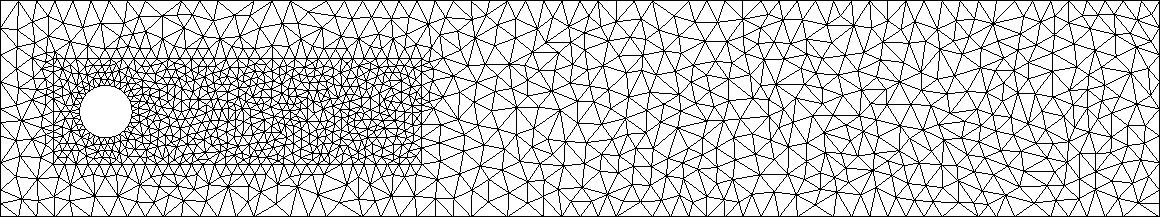};

\path [draw=black, fill opacity=0] (axis cs:13,1)--(axis cs:13,1);

\path [draw=black, fill opacity=0] (axis cs:0.05,13)--(axis cs:0.05,13);

\path [draw=black, fill opacity=0] (axis cs:13,1.38777878078145e-17)--(axis cs:13,1.38777878078145e-17);

\path [draw=black, fill opacity=0] (axis cs:0,13)--(axis cs:0,13);

\node at (axis cs:0.,0.25) [color=black, fill=white, anchor=south west] {$\Gamma_0$};
\node at (axis cs:2.2,0.21) [color=black, fill=white, anchor=south east] {$\Gamma_1$};

\end{axis}

\end{tikzpicture}

%% file: pics/cylwake.tikz
% This file was created by matplotlib v0.1.0.
% Copyright (c) 2010--2014, Nico Schlömer <nico.schloemer@gmail.com>
% All rights reserved.
% 
% The lastest updates can be retrieved from
% 
% https://github.com/nschloe/matplotlib2tikz
% 
% where you can also submit bug reports and leavecomments.
% 
\begin{tikzpicture}

\begin{axis}[
xmin=0, xmax=2.2,
ymin=0, ymax=.41,
width=\figurewidth,
height=\figureheight,
axis on top,
xtick={0,0.2,0.32, 0.6, 0.7, 2.2},
ytick={0,.15,.25,.41},
tick label style={font=\footnotesize},
xlabel={$x_0$},
ylabel={$x_1$},
xlabel near ticks,
ylabel near ticks
]
\addplot graphics [includegraphics cmd=\pgfimage,xmin=0, xmax=2.2, ymin=0, ymax=0.41] {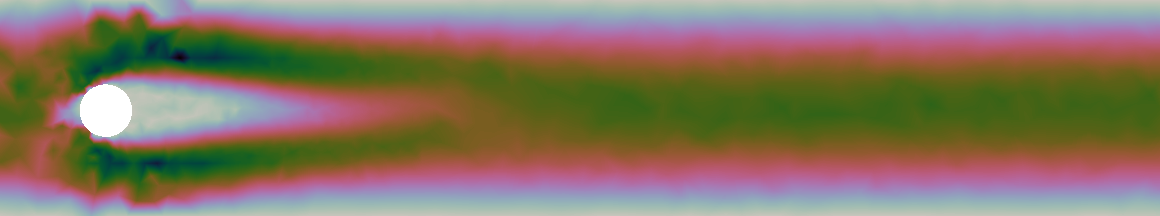};

\path [draw=black, fill opacity=0] (axis cs:13,1)--(axis cs:13,1);

\path [draw=black, fill opacity=0] (axis cs:0.05,13)--(axis cs:0.05,13);

\path [draw=black, fill opacity=0] (axis cs:13,1.38777878078145e-17)--(axis cs:13,1.38777878078145e-17);

\path [draw=black, fill opacity=0] (axis cs:0,13)--(axis cs:0,13);

\draw [color=black] (axis cs:0.27,0.15) rectangle (axis cs:0.32,0.25);
\node at (axis cs:0.3,0.23) [color=black, anchor=north west] {$\Omega_c$};
\draw [color=white] (axis cs:0.6,0.15) rectangle (axis cs:0.7,0.25);
\node at (axis cs:0.68,0.24) [color=white, anchor=north west] {$\Omega_o$};
\end{axis}

\end{tikzpicture}

%% file: controlsetups.tex
\section{Modelling and Approximation of Control and Observation}\label{sec:modappcont}
In flow control, one tries to act on the flow in such a way that certain flow pattern are stabilized or enforced by means of actuators. Also, sensors are used to monitor the process and to provide necessary information for the actuation. For the simulation, actuation and measuring are modelled through input and output operators $B$ and $C$ and corresponding spaces for the input and output signals $u$ and $y$ that are included in the equations as follows:

\begin{subequations}\label{eq:NSio}
\begin{align}
  \frac{\partial v}{\partial t}+(v\cdot\nabla)v-\Rei\Delta v+\nabla p&=f+Bu,\\
		\nabla\cdot v&=0, \\
		y &= C(v,p).
\end{align}
\end{subequations}

Distributed control, where the control acts as a volume force distributed in a domain of control, is hardly realizable in practice. Nevertheless, we provide models for distributed control, because the modelling and spatial discretization naturally leads to a standard state-space system, which is the basis for typical control or model reduction algorithms.

Boundary control through Dirichlet boundary conditions for the velocity, which is the setup for many flow control problems, does not simply lead to a standard state space representation, cf. \cite{BenH15a}. We provide boundary control models in state space form by means of Robin approximations of the Dirichlet boundary conditions.

\subsection{Distributed Control and Observation}

We define input and output state spaces $U$ and $Y$ as subspaces of $L^2([0,1])$.

Let $\Omega$ be the computational domain and $\Omega_c=[x^c_{0,min}, x^c_{0,max}]\times[x^c_{1,min}, x^c_{1,max}]$ a rectangular subdomain where the control acts. We define the input operator $B\colon U\times U \to [L^2(\Omega)]^2$ such that the control acts distributed in the domain of control $\Omega_c$ with two spatial components.

In order to save some degrees of freedom, we assume that $\Bop u$ is constant in one spatial direction -- say $x_0$.

Concretely, for $\mathcal U:=L^2(0,T;U\times U)$ we define the $\Bop \colon \mathcal U \to  L^2(0,T; L^2(\Omega;\mathbb R^{2} ))$ via 
\begin{equation}\label{eq:defInputOp}
	\Bop u(t;x_0,x_1)=
\begin{cases}
			u(t;\theta(x_1)) , & \quad\mbox{for $(x_0,x_1)\in \Omega_c$,} \\
		[0, 0]^T, &\quad\mbox{elsewhere,}
	\end{cases} 
\end{equation}
for $u\in \mathcal U$ and with the affine linear function $\theta_c$ mapping the $x_1$-extension of $\Omega_c$, $[x^c_{1,min}, x^c_{1,max}]$, onto $[0, 1]$. Note that the roles of $x_0$ and $x_1$ can be interchanged depending on the setup.

The output operator $\mathcal P_Y \Cv \colon L^2(0,T; L^2(\Omega; \mathbb R^{s}) \to L^2(0,T; Y\times Y)$ for the velocity is defined as follows
\begin{enumerate}
	\item For $v(t)\in L^2(\Omega, \mathbb R^{2} )$ compute $\Cv v(t) \in L^2(0,1; \mathbb R^2)$ through 
\begin{equation}\label{eq:defOutputOp}
	\Cv v(t)(\eta) = 
		\tfrac 1{x^o_{0,max}-x^o_{0,min}}\int_{x^o_{0,min}}^{x^o_{0,max}} 
		v(t;x_0,\theta_o(\eta))
		\inva {x_0},
\end{equation}
		which, basically, is the velocity in the domain of observation $\Omega_o$ averaged in $x_0$ direction. Here, $\theta_o$ is an affine linear mapping adjusting $[0,1]$ to $[x^o_{1,min}, x^o_{1,max}]$.
	\item Define the corresponding discrete signal $y(t)\in L^2(0,T;Y\times Y)$ as $y(t) = \mathcal P_{\mathcal Y} \Cv v(t)$, where $\mathcal P_{Y}\colon [L^2(0,1)]^2 \to Y\times Y$ is the $L^2$-orthogonal projection.
\end{enumerate}

Namely, for a rectangular domain of observation $\Omega_o = [x^o_{0,min}, x^o_{0,max}]\times[x^o_{1,min}, x^o_{1,max}]$, for $\mathcal Y:= L^2(0,T; Y\times Y)$, and for a $v\in L^2(0,T;L^2(\Omega;\mathbb R^{2} ))$, we define the observation operator $\Cv\colon v \mapsto y \in \mathcal Y$ via

As a pressure-based output we take the average pressure in a subdomain $$\Omega_p = [x^p_{0,min}, x^p_{0,max}]\times[x^p_{1,min}, x^p_{1,max}],$$
i.e. 
\begin{equation}\label{eq:output_def}
	y_p(t) = \Cp p(t) := \frac{1}{|\Omega_p|}\int_{\Omega_p} p(t,x) dx
\end{equation}
evaluated in the corresponding finite element space.

\subsection{Assembling of the Distributed Control and Observation Operators}
To setup a discrete representation of the input operator $B$, we impose several assumptions on the space-time structure of the input signals. 

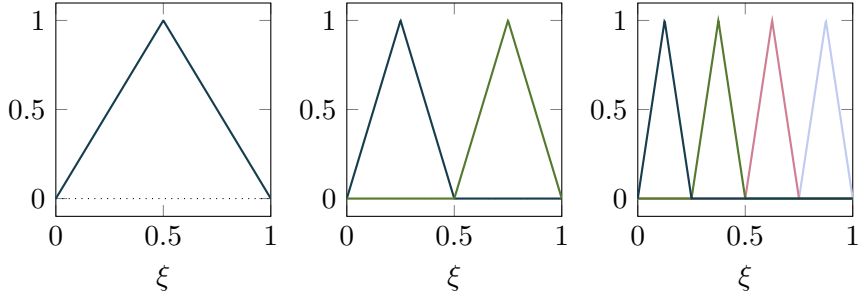
\begin{figure}[ht]
 \centering
 \input{pics/basfuncs.tikz}
\caption{Three levels of a hierarchical basis for piecewise linear functions for nested $L^2(0,1)$-subspaces.}
\label{fig:hierarchical}
\end{figure}

Firstly, for a given $\Nu \in 2\mathbb N^{}$, we choose the signal state space $U\times U$ as $U_0 \oplus U_1$, where 
$U_0 := \spann\{ \vec\nu_l \}_{l=1, \dotsc, \Nu/2}$ and $U_1:= \spann \{ \vec\nu_l \}_{l=\Nu/2+1, \dotsc, \Nu}$, where 
\begin{equation}\label{eq:io_vec_basfuncs}
	\vec\nu_l := \bbmat \nu_l \\ 0 \ebmat, \quad \text{for }l=1,\dotsc, \Nu/2 
	\andi
	\vec\nu_l := \bbmat 0 \\ \nu_l  \ebmat, \quad \text{for }l=\Nu/2+1 , \dotsc, \Nu,
\end{equation}
with $\nu_l$ being the hierarchical piecewise linear hat functions illustrated in Figure \ref{fig:hierarchical}.

Note that by defining $U_0$ and $U_1$ with only one spatial coordinate $\xi$, we anticipate that the input is to be defined depending only on one spatial coordinate, cf. \eqref{eq:defInputOp}, and note that bases of subspaces of $L^2(0,1)$ other than shown in Figure \ref{fig:hierarchical} can be used.

Secondly, we assume that the time and space component in the chosen discrete input space are separated, i.e., that we can write the inputs as 
\begin{equation}
	u(t;\xi) = \sum_{l=1}^{\Nu}s_l(t)\vec \nu_l(\xi) . \label{eq:def_u_sprtd}
\end{equation}

With $u$ defined as \eqref{eq:def_u_sprtd} and the input to the system $\Bop u$ as defined in \eqref{eq:defInputOp} discretized like the right hand side as in \eqref{eq:coeffs_massrhs}, the discrete input operator $\Bd\in \mathbb R^{\Nu}\to \mathbb R^{\Nv}$, that maps the vector $\mathbf u = [s_1, \dotsc, s_{\Nu}]^T$ of control coefficients, cf. \eqref{eq:def_u_sprtd}, onto the control action $\Bd \mathbf u$ is defined as 

\begin{equation}\label{eq:defdiscB}
	\Bd=\left[ \int_\Omega \phi_i: \Bop \vec \nu_l  \inva x \right]_{i=1,\cdots, \Nv;~l=1,\cdots,\Nu}.
\end{equation}

For the output space for the velocity measurements, given $\Ny \in 2\mathbb N$, we define $Y\times Y = Y_0 \oplus Y_1$ similar to $U \times U$ with basis functions for both spatial components as in \eqref{eq:io_vec_basfuncs}. As basis functions we use $\Ny$ piecewise linear hat functions defined on $[0, 1]$ partitioned into $\Ny/2-1$ equidistant segments. From the space-time separated ansatz for the discrete velocity \eqref{eq:vpexpansion} and from the definition of the output operator in \eqref{eq:output_def} it follows that the measured output will always be of the form
$$\mathcal P_Y \Cv v(t)(\eta) =\sum_{j=1}^{\Nv}v_j(t)\mathcal P_Y\Cv \phi_j(\eta).
$$
Since for every $j=1,\cdots, \Nv$, the term $\mathcal P_{\mathcal Y}\Cv \phi_j$ is in $Y\times Y$ and, thus, can be expanded in the basis \eqref{eq:io_vec_basfuncs}, the measurement from the discrete velocity can be written as 
\begin{equation} \label{eq:yexpand}
	\mathcal P_Y\Cv v(t)(\eta) = \sum_{l=1}^{\Ny} y_k(t) \vec \nu_l(\eta),
\end{equation}
with a unique vector of coefficients $\mathbf y := [y_1, \cdots , y_\Ny]^T$.

These coefficients are obtained through the discrete approximation $\Cdv$ to $\mathcal P_Y \Cv$ that maps the coefficient vector $\mathbf v$ of the velocity approximation onto $\mathbf y$ and that is defined through

\begin{equation}
	\Cdv =\My^{-1}  \left[\int_0^1 \vec \nu_l(\eta) : \Cv \phi_j(\eta) \inva \eta \right]_{l=1,\cdots,\Ny;~j=1,\cdots,\Nv}.
	\label{eq:defdiscC}
\end{equation}
This assembling is basically taking the output for all basis functions, testing them against the basis functions of $Y\times Y$, and premultiplying the obtained matrix by the inverse of the mass matrix $\My := \left[ \int_0^1 \vec \nu_l(\eta) : \vec \nu_k(\eta) \inva \eta \right]_{k,l = 1,\cdots,\Ny}$. 

The pressure based output as defined in \eqref{eq:output_def} only depends on time, i.e. $\Cp\colon L^2(0,T; \mathbb R^{\Np}) \to L^2(0,T)$. Thus, the discrete representation $\Cdp$ of $\Cp$ that maps the coefficient vector $\mathbf p$ of a discrete pressure $p$, cf. \eqref{eq:vpexpansion}, onto $\Cp p \in L^2(0,T)$ is defined through
\begin{equation} \label{eq:defdiscCp}
	\Cdp = \left[ \Cp \psi_k \right]_{;k=1,\cdots,\Np}.
\end{equation}

Depending on the setup, in the provided data files, the discrete input and output operators appear as listed in Table \ref{tab:ioparams}. In \lstinline$B$, the first $\Nu/2$ columns are associated with the control in $x_0$ direction and the last $\Nu/2$ columns with control in $x_1$ direction. Thus, to vary the number of control parameters one may discard a control direction or, thanks to the hierarchy of the basis functions, only use, e.g., the first three columns. Similarly, in \lstinline$C$ the first $\Ny/2$ rows correspond to the $x_0$ component of the output, while the last $\Ny/2$ rows correspond to the $x_1$ component. To reduce the output dimension, one may also discard one component. Also one can discard single degrees of freedom in the output space. Note, however, that the basis in the output space is not hierarchical but the standard hat function basis.
\begin{table}[tb]
	\centering
	\begin{tabular}{lcl}
		\lstinline$B$ & $\leftrightarrow$ & $[\Bd]_I$\\
		\lstinline$Cv$ & $\leftrightarrow$ & $[\Cdv]_I$\\
		\lstinline$Cp$ & $\leftrightarrow$ & $\Cdp$\\
	\end{tabular}
	\quad and \quad
	\begin{tabular}{lcl}
		\lstinline$My$& $\leftrightarrow$ &$\mathcal M_Y$\\
		\lstinline$Mu$& $\leftrightarrow$ &$\mathcal M_U$\\
	\end{tabular}
	\caption{Parameters defining the discrete input and output operators and the mass matrices of the signal state spaces.}
	\label{tab:ioparams}
\end{table}

\subsection{Boundary Control}
A common case of boundary control of a flow is defining a time-varying velocity profile on a part of the boundary. This models, for example, a valve at the peripherie where one can inject a jet into the flow or exhaust certain amounts of the fluid.

For illustration, we assume that such a control takes places at a single part $\Gamma_c$ of the boundary $\Gamma$ and that we have constant Dirichlet or \emph{do-nothing} conditions elsewhere. The generalization to more controlled segments of the boundary and to other boundary conditions elsewhere is straight forward.

So assume that the control action is given as 
\begin{equation}\label{eq:diricont}
	v(t)\bigr|_{\Gamma_c} = u(t).
\end{equation}
% We will not address how relation \eqref{eq:diricont} has to be defined in order to make sense and in order to admit a solution to the Navier-Stokes equations \eqref{eq:NSintro}, but refer the reader to the discussion in Section \ref{sec:discnrefs}.

One may enforce the Dirichlet conditions by simply setting the values of the discrete approximations at the boundary. However, this leads to the appearance of $\dot u$, cf. \cite{BenH15a}. Therefore, we relax the Dirichlet to Robin conditions and approximate \eqref{eq:diricont} via
\begin{equation}\label{eq:dirirelrobcont}
	u(t) \approx v(t)\bigr|_{\Gamma_c} + \alpha (p \vec n - \nu \frac{\partial v}{\partial \vec n})\bigr|_{\Gamma_c},
\end{equation}
with a penalization parameter $\alpha$ that is intended to go to zero. 
This approximation of the Dirichlet control has been investigated in the context of optimal control of Navier--Stokes equations in \cite{HouR99}.

\subsection{Assembling of the Boundary Control Operator}\label{sec:ass_bcc_op}
We assume that $u$ is given as a sum of spatial shape functions with time dependent coefficients, i.e. $u$ has the form $u(x,t)=\sum_{l=1}^{\Nu}g_l(x)u_l(t)$. 

Rearranging \eqref{eq:dirirelrobcont} to $p \vec n - \nu \frac{\partial v}{\partial \vec n}\bigr|_{\Gamma_c} \approx \frac 1\alpha u(t) - \frac 1\alpha v(t)\bigr|_{\Gamma_c}$, the control is numerically realized through the term $f_N$, cf. \eqref{eq:normalbcsterm}, which gives the linear coefficient
\begin{equation}\label{eq:def_a_rbc}
	\Abc = \frac 1\alpha \left[ \int_{\Gamma_N} \phi_i : \phi_j \inva x \right]_{i,j =1,\dotsc,\Nv},
\end{equation}
and the source term, $\Bbc\mathbf u$, where $\mathbf u:=[u_1, \dotsc, u_\Nu]^T$ and
\begin{equation*}
	\Bbc = \frac 1\alpha \left[ \int_{\Gamma_N} \phi_i : g_l \inva x \right]_{i =1,\dotsc,\Nv;~ l=1,\dotsc,\Nu}.
\end{equation*}

Depending on the setup, in the provided data files, the discrete boundary control operators appear as listed in Table \ref{tab:bcparams}. There also $\alpha$ is set to $1$ so that different penalization parameter values can be realized through scaling \lstinline$Abc$ and \lstinline$Bbc$ correspondingly.
\begin{table}[tb]
	\centering
	\begin{tabular}{lcl}
		\lstinline$Bbc$ & $\leftrightarrow$ & $[\Bbc]_I$\\
		\lstinline$Abc$ & $\leftrightarrow$ & $[\Abc]_I$\\
	\end{tabular}
	\caption{The linear operators realizing the boundary control through the Robin relaxation with $\alpha=1$.}
	\label{tab:bcparams}
\end{table}
% Note the following remarks concerning the interference of Dirichlet boundary control via Robin approximations with the remaining boundary conditions.

	Note, that there is no constant contribution of other Dirichlet boundary values through $\Abc \bbug$, cf. \eqref{eq:resobcnodes}. A contribution may occur where a Dirichlet and a Robin node share some support on $\Gamma_N$ (cf. \eqref{eq:def_a_rbc}) which can only happen where Dirichlet and control boundaries meet. However, in the considered setups the control boundaries are in the neighborhood of zero Dirichlet so that the possible contribution would be zeroed out by the prescribed node value.

	If one considers Robin control, the set $\cI\sI$, cf. \eqref{eq:innerbcnodes}, will also contain the Robin nodes as if they were inner nodes. Accordingly, these nodes are also solved for in the process of the simulations and they simply integrated as inner nodes in the assembling of the linear and nonlinear operators; cf. Sections \ref{sec:asslinop} and \ref{sec:assnonlinop}.

	\section{Control Setups}
We describe the particular setups that include control and observation.

\subsection{Driven Cavity with Distributed Control and Observation}\label{sec:drivcavdistcont}
We consider the test case of the lid driven cavity described in Section \ref{sec:drivcavplain} and add distributed control and observation that are defined and assembled as described in Section \ref{sec:modappcont}.

Specifically, we define the domain of control and observation of the velocity as $\Omega_c:=[0.4,0.6]\times [ 0.2,0.3]$ and $\Omega_o := [0.45,0.55] \times [0.5,0.7]$, see Figure \ref{fig:drivcav}. The pressure is observed in $\Omega_p := [0.45,0.55] \times [0.7,0.8]$.

Recall that the input and the velocity output were modelled with only one dimensional space dependencies, cf. \eqref{eq:defInputOp} and \eqref{eq:defOutputOp}. For the driven cavity the spatial components of the signal are modelled as follows

\begin{table}[h]
		\centering
		\begin{tabular}{c|c|c}
			& $x_0$-direction& $x_1$-direction \\
			\hline
			input & space-varying & constant \\
			output & constant (averaged) & space-varying
		\end{tabular}
		\caption{Spatial components of the input and velocity output signals for the distributed control of the driven cavity problem.}
		\label{tab:ssigdrivcav}
	\end{table}

% The input is modelled to be constant in space in $x_1$ direction and space varying in $x_0$ direction. The velocity output is taken as space varying in $x_1$ direction and averaged in $x_0$ direction.
\subsection{Cylinder Wake with Distributed Control and Observation}\label{sec:cylwakedistr}
The test case of the cylinder wake is described in Section \ref{sec:cylwakeplain}. Again, we add distributed control and observation as defined in Section \ref{sec:modappcont}.

Specifically, we define the domain of control and observation of the velocity as $\Omega_c:=[0.27,0.32]\times [ 0.15,0.25]$ and $\Omega_o := [0.6,0.7] \times [0.15,0.25]$, see Figure \ref{fig:cylwake}. The pressure is observed in $\Omega_p := [0.6,0.64] \times [0.18,0.22]$.

Recall that the input and the velocity output were modelled with only one dimensional space dependencies, cf. \eqref{eq:defInputOp} and \eqref{eq:defOutputOp}. For the cylinder wake, the spatial components of the signal are modelled as follows

\begin{table}[h]
		\centering
		\begin{tabular}{c|c|c}
			& $x_0$-direction& $x_1$-direction \\
			\hline
			input & constant & space-varying \\
			output & constant (averaged) & space-varying
		\end{tabular}
		\caption{Spatial components of the input and velocity output signals for the distributed control of the driven cavity problem.}
		\label{tab:ssigcylwake}
	\end{table}
\subsection{Cylinder Wake with Boundary Control}\label{sec:cylwakebccont}
The setup of the simulation and of the output definition is the same as described in Section \ref{sec:cylwakedistr}. For the input we define and assemble a penalized Robin scheme as explained in Section \ref{sec:ass_bcc_op}. 
Specifically, we define the control through injection and suction at outlets $\Gamma_{c_1}$, $\Gamma_{c_2}$ centered at the cylinder periphery at $\pm \pi/3$ occupying $\pi/6$ of the circumference.
		Then, we prescribe Dirichlet conditions for the velocity
			$$v=n_1g_1(x)u_1(t), \quad v=n_2g_2(x)u_2(t),$$
			at $\Gamma_{c_1}$ and $\Gamma_{c_2}$, where $n_{1/2}$ are the vectors that are normal to the outlets in their center and that point into the domain, where $u_{1/2}$ are the magnitudes of the controls, and where $g_{1/2}$ are the shape functions that parametrize the segments $\gaco /\gact$ via $s\in[0, 1]$ and that return the value of 
 $$1 - 0.5(1 + \sin( (2s + 0.5)\pi)), $$
 thus realizing a shape function that is $1$ in the center of the control outlets and that smoothly extends to the zero \emph{no-slip} conditions at the adjacent boundaries.
	\begin{figure}
		{\includegraphics[scale=0.6]{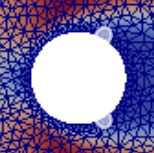}}
		\caption{Illustration of the control locations on the cylinder periphery.}
	\end{figure}

%% file: pics/basfuncs.tikz
\begin{tikzpicture} 
\begin{groupplot}[group style={group size=3 by 1},width=0.3\textwidth,height=0.3\textwidth] 
\nextgroupplot[xmin=0, xmax=1, xtick={0.0,0.5,1.0}, xlabel=$\xi$]
\addplot[color=black,dotted] coordinates {(0,0) (1,0)};
\addplot[color=color0,thick] table [x = {x}, y = {mu3}] {pics/PLF.dat};
\nextgroupplot[xmin=0, xmax=1, xtick={0.0,0.5,1.0}, xlabel=$\xi$] 
\addplot[color=black,dotted] coordinates {(0,0) (1,0)};
\addplot[color=color0,thick] table [x = {x}, y = {mu4}] {pics/PLF.dat};
\addplot[color=color1,thick] table [x = {x}, y = {mu5}] {pics/PLF.dat};
\nextgroupplot[xmin=0, xmax=1, xtick={0.0,0.5,1.0}, xlabel=$\xi$] 
\addplot[color=black,dotted] coordinates {(0,0) (1,0)};
\addplot[color=color3,thick] table [x = {x}, y = {mu9}] {pics/PLF.dat};
\addplot[color=color2,thick] table [x = {x}, y = {mu8}] {pics/PLF.dat};
\addplot[color=color1,thick] table [x = {x}, y = {mu7}] {pics/PLF.dat};
\addplot[color=color0,thick] table [x = {x}, y = {mu6}] {pics/PLF.dat};
\end{groupplot} 
\end{tikzpicture}

%% file: numexa.tex
\section{The Data, Plotting, and Auxiliary Functions}
\subsection{The System Matrices Files}
The provided data, see Table \ref{fig:linkcodndat}, are vectors and sparse matrices stored as \lstinline$.mat$ files in the \emph{MATLAB 6} version. 

The naming of the data is as
$$
\mlsti{PROBLEMNAME__mats__NV*XX*_Re1[_bccontrol_palpha1].mat}
$$
where
\begin{itemize}
  \item \lstinline$PROBLEMNAME={cylinderwake,drivencavity}$,
  \item \lstinline$*XX*$ denotes the size of the system, namely the dimension of the velocity approximation space, and 
  \item \lstinline$[_bccontrol_palpha1]$ is present, if boundary control is realized.
\end{itemize}

\subsection{Files for Visualizations}
Visualization is done through writing the values of the pressure or velocity approximation to \lstinline$.vtu$ files that can be viewed in \emph{Paraview}. For this there are \emph{json} files that, among others, contains the geometrical information of the mesh and the map of the coordinates of the approximating vectors to the mesh vertices and that are named as
$$
  \mlsti{visualization_PROBLEMNAME_N*XX*.jsn}
$$
with \lstinline$PROBLEMNAME$ as above and \lstinline$*XX*$ standing for the discretization level. Note that \lstinline$N$ is different from \lstinline$NV$ used for the naming of the system matrices, since on a given mesh topology one can define approximations of, e.g., different degrees of freedom. For the provided data, the relations between mesh parameters \lstinline$N$ and approximation scheme parameters \lstinline$NV$ are listed in Table \ref{tab:meshparams}.

\begin{table}[tb]
	\centering
	\begin{tabular}{l|l|l}
		\lstinline$drivencavity$ & \lstinline$cylinderwake$ & \lstinline$cylinderwake_bccontrol$\\
		\hline
		\lstinline$N=10$ $\leftrightarrow$ \lstinline$NV=722$
		&
		\lstinline$N=1$ $\leftrightarrow$ \lstinline$NV=5812$
		&
		\lstinline$N=1$ $\leftrightarrow$ \lstinline$NV=5824$
		\\
		\lstinline$N=20$ $\leftrightarrow$ \lstinline$NV=3042$
		&
		\lstinline$N=2$ $\leftrightarrow$ \lstinline$NV=9356$
		&
		\lstinline$N=2$ $\leftrightarrow$ \lstinline$NV=9384$
		\\
		\lstinline$N=30$ $\leftrightarrow$ \lstinline$NV=6962$
		&
		\lstinline$N=3$ $\leftrightarrow$ \lstinline$NV=19468$
		&
		\lstinline$N=3$ $\leftrightarrow$ \lstinline$NV=19512$
		\\
	\end{tabular}
	\caption{Relation of the mesh parameters \lstinline$N$ and the discretization parameter \lstinline$NV$ for the considered problems and the provided data.}
	\label{tab:meshparams}
\end{table}

\subsection{Auxiliary Files}
The data comes with two software modules, namely \lstinline$conv_tensor_utils$ that provides functionality for linearization and memory-efficient evaluation of the nonlinearity and \lstinline$visualization_utils$ that assists in writing output files for rendering in \emph{Paraview}.

\section{Numerical Examples}
We illustrate the use of the data in some example setups.
\subsection{Cylinder Wake}
In this section, we consider example setups for the cylinder wake as described in Sections \ref{sec:cylwakeplain}, \ref{sec:cylwakedistr}, and \ref{sec:cylwakebccont}.
\subsubsection{Steady State with \lstinline$Re=40$}\label{ssec_cylwake_stst_Re40}
\fbox{
\begin{minipage}[t]{.9\textwidth}
\begin{minipage}[t]{.4\textwidth}
	\begin{tabular}{l}
	  Setup: \hfill \lstinline$cylinderwake$ \\
   Steady state simulation\\
   Reynolds number: \hfill \lstinline$Re=40$\\
   Grid: \hfill \lstinline$N=1$
 \end{tabular}
\end{minipage}
\hfill
\begin{minipage}[t]{.4\textwidth}
	\begin{tabular}{l}
  Simulation results: \\
  \lstinline$|v|$: \hfill Figure \ref{fig_cylwake_stst_Re40}\\
  \lstinline$p$: \hfill Figure \ref{fig_cylwake_stst_Re40}
 \end{tabular}
\end{minipage}
\\[.15in]
To replicate this simulation run
$$
  \mlsti{cylinderwake_steadystate.\{m,py\}}
$$
\end{minipage}
}

\begin{figure}[p]
  {\includegraphics[scale=0.35]{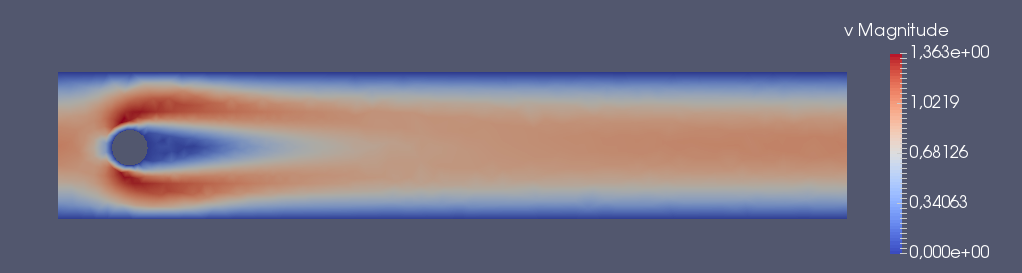}}
  {\includegraphics[scale=0.351]{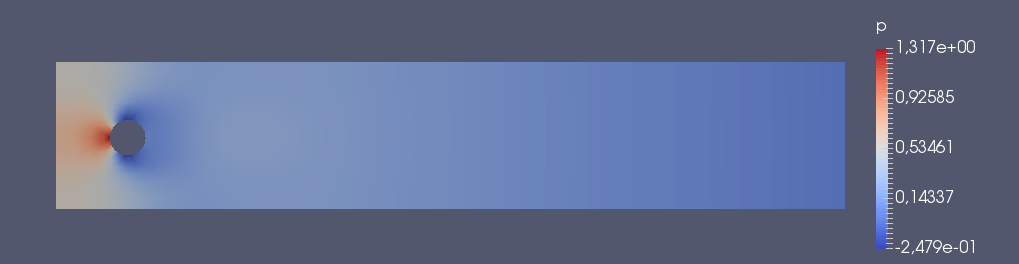}}
  \caption{Illustration of the velocity magnitude and pressure field for the setup described in Section \ref{ssec_cylwake_stst_Re40}}
  \label{fig_cylwake_stst_Re40}
\end{figure}

\subsubsection{Steady State with \lstinline$Re=40$ with Boundary Control Impact}\label{ssec_cylwake_stst_Re40_bccont}
\fbox{
\begin{minipage}[t]{.9\textwidth}
\begin{minipage}[t]{.4\textwidth}
	\begin{tabular}{l}
	  Setup: \hfill \lstinline$cylinderwake$ \\
   Steady state simulation\\
   Reynolds number: \hfill \lstinline$Re=40$\\
   Grid: \hfill \lstinline$N=2$\\
   Robin-Penalization: \hfill \lstinline$palpha=1e-3$\\
   Input: \hfill \lstinline$uvec=[1,-1].T$
 \end{tabular}
\end{minipage}
\hfill
\begin{minipage}[t]{.4\textwidth}
	\begin{tabular}{l}
  Simulation results: \\
  \lstinline$|v|$: \hfill Figure \ref{fig_cylwake_stst_Re40_bccont}\\
  \lstinline$p$: \hfill Figure \ref{fig_cylwake_stst_Re40_bccont}
 \end{tabular}
\end{minipage}
\\[.15in]
To replicate this simulation run
$$
  \mlsti{cylinderwake_steadystate_bccontrol.\{m,py\}}
$$
\end{minipage}
}

\begin{figure}[p]
  {\includegraphics[scale=0.35]{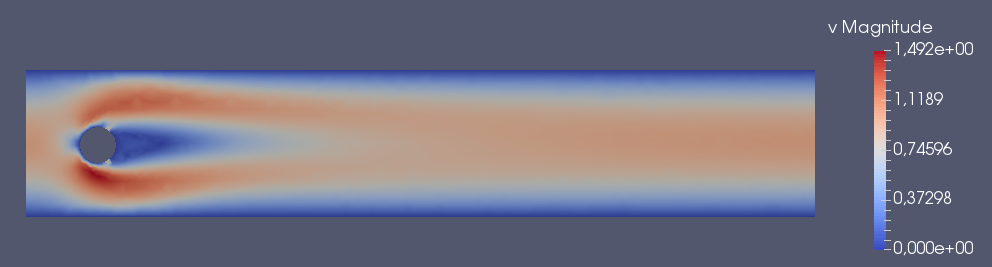}}
  {\includegraphics[scale=0.351]{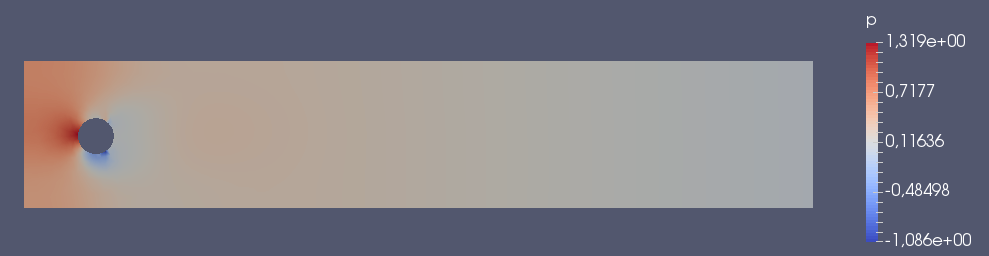}}
  \caption{Illustration of the velocity magnitude and pressure field for the setup described in Section \ref{ssec_cylwake_stst_Re40_bccont}}
  \label{fig_cylwake_stst_Re40_bccont}
\end{figure}

\subsubsection{Transient Case with \lstinline$Re=90$}\label{ssec_cylwake_tdp_Re90}
\fbox{
\begin{minipage}[t]{.9\textwidth}
\begin{minipage}[t]{.4\textwidth}
	\begin{tabular}{l}
	  Setup: \hfill \lstinline$cylinderwake$ \\
   Transient simulation with IMEX-Euler\\
   Reynolds number: \hfill \lstinline$Re=90$\\
   Grid: \hfill \lstinline$N=3$\\
   Time grid: \lstinline$t0, tE, Nts = 0., 4., 2**11$ \\
	 Initial value: \hfill \emph{Stokes solution}
 \end{tabular}
\end{minipage}
\hfill
\begin{minipage}[t]{.4\textwidth}
	\begin{tabular}{l}
  Simulation results: \\
  \lstinline$|v[t=tE]|$: \hfill Figure \ref{fig_cylwake_tdp_Re90_t4}\\
  \lstinline$p[t=tE]$: \hfill Figure \ref{fig_cylwake_tdp_Re90_t4}\\
  \lstinline$C*v$: \hfill Figure \ref{fig_cylwake_tdp_Re90_voutpout}\\
  \lstinline$C*p$: \hfill Figure \ref{fig_cylwake_tdp_Re90_voutpout}
 \end{tabular}
\end{minipage}
\\[.15in]
To replicate this simulation run
$$
  \mlsti{cylinderwake_tdp_vout_pout.\{m,py\}}
$$
\end{minipage}
}

\begin{figure}[p]
  {\includegraphics[scale=0.3528]{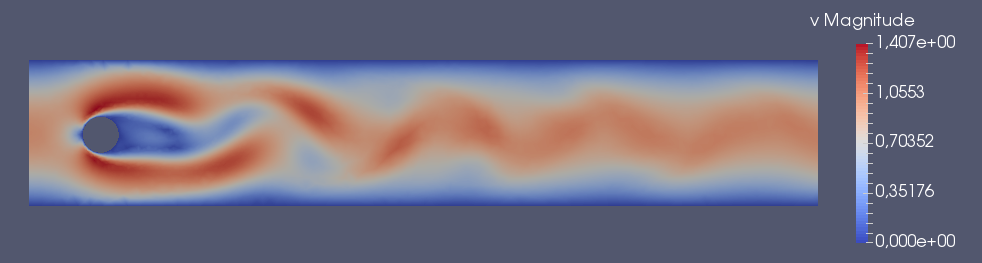}}
  {\includegraphics[scale=0.35]{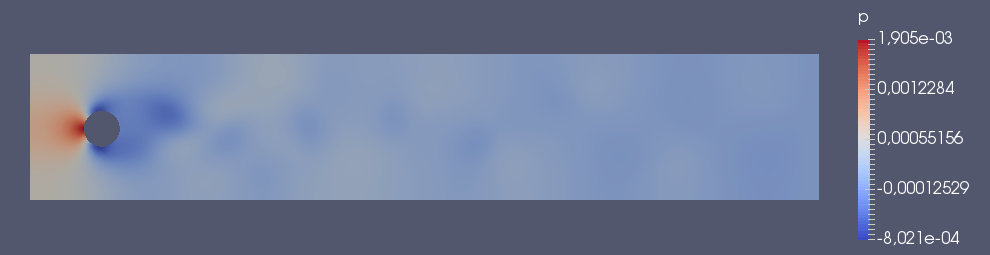}}
  \caption{Illustration of the velocity magnitude and pressure field at \lstinline$t=4.0$ for the setup described in Section \ref{ssec_cylwake_tdp_Re90}}
  \label{fig_cylwake_tdp_Re90_t4}
\end{figure}

\setlength\figurewidth{6cm}
\setlength\figureheight{5cm}
\begin{figure}[p]
\input{plots/v_nsequadtens-N3-tE4.0-Nts2048.tikz}
\input{plots/p_nsequadtens-N3-tE4.0-Nts2048.tikz}
  \caption{Illustration of the velocity and pressure output for the setup described in Section \ref{ssec_cylwake_tdp_Re90}}
  \label{fig_cylwake_tdp_Re90_voutpout}
\end{figure}

\subsubsection{Transient Case with \lstinline$Re=40$ and Boundary Actuation}\label{ssec_cylwake_tdp_Re40_bccont}
\fbox{
\begin{minipage}[t]{.9\textwidth}
\begin{minipage}[t]{.6\textwidth}
	\begin{tabular}{l}
	  Setup: \hfill \lstinline$cylinderwake$ \\
   Transient simulation with IMEX-Euler\\
   Reynolds number: \hfill \lstinline$Re=40$\\
   Grid: \hfill \lstinline$N=2$\\
   Time grid: \hfill \lstinline$t0, tE, Nts = 0., 6., 2**10$\\
	 Initial value: \hfill \emph{Stokes solution} \\
   Robin-Penalization: \hfill \lstinline$palpha=1e-3$\\
   Input frequency:\hfill \lstinline$omega=3*Pi/(tE-t0)$ \\ 
   Input: \hfill \lstinline$uvec=sin(omega*t)*[1,-1].T$
 \end{tabular}
\end{minipage}
\hfill
\begin{minipage}[t]{.3\textwidth}
	\begin{tabular}{l}
  Simulation results: \\
  \lstinline$C*v$: \hfill Figure \ref{fig_cylwake_tdp_Re40_bccont_voutpout}\\
  \lstinline$C*p$: \hfill Figure \ref{fig_cylwake_tdp_Re40_bccont_voutpout}
 \end{tabular}
\end{minipage}
\\[.15in]
To replicate this simulation run
$$
  \mlsti{cylinderwake_tdp_pout_vout_bccontrol.\{m,py\}}
$$
\end{minipage}
}

\setlength\figurewidth{6cm}
\setlength\figureheight{5cm}
\begin{figure}[p]
\input{plots/v_nsequadtens-N2-tE6.0-Nts1024-bccomg3.0.tikz}
\input{plots/p_nsequadtns-N2-tE6.0-Nts1024-bccomg3.0.tikz}
  \caption{Illustration of the velocity and pressure output for the setup described in Section \ref{ssec_cylwake_tdp_Re40_bccont}}
  \label{fig_cylwake_tdp_Re40_bccont_voutpout}
\end{figure}

\subsection{Driven Cavity}
In this section, we consider example setups for the driven cavity as described in Sections \ref{sec:drivcavplain} and \ref{sec:drivcavdistcont}.
\subsubsection{Steady State with \lstinline$Re=1200$}\label{ssec_drivcav_stst_Re1200}

\fbox{
\begin{minipage}[t]{.9\textwidth}
\begin{minipage}[t]{.4\textwidth}
	\begin{tabular}{l}
	  Setup: \hfill \lstinline$drivencavity$ \\
   Steady state simulation\\
   Reynolds number: \hfill \lstinline$Re=1200$\\
   Grid: \hfill \lstinline$N=30$
 \end{tabular}
\end{minipage}
\hfill
\begin{minipage}[t]{.4\textwidth}
	\begin{tabular}{l}
  Simulation results: \\
  \lstinline$|v|$: \hfill Figure \ref{fig_drivcav_stst_Re1200}\\
  \lstinline$p$: \hfill Figure \ref{fig_drivcav_stst_Re1200}
 \end{tabular}
\end{minipage}
\\[.15in]
To replicate this simulation run
$$
  \mlsti{drivencavity_steadystate.\{m,py\}}
$$
\end{minipage}
}

\begin{figure}[p]
	{\includegraphics[scale=0.345]{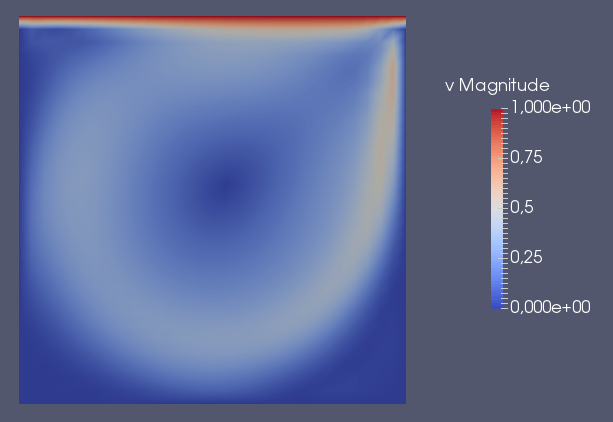}}
	{\includegraphics[scale=0.35]{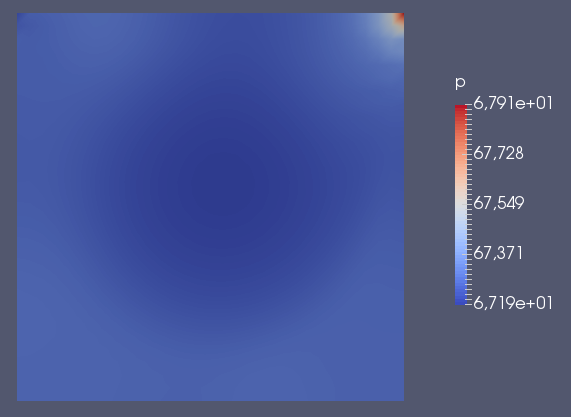}}
	\caption{Illustration of the velocity magnitude and pressure field for the setup described in Section \ref{ssec_drivcav_stst_Re1200}}
  \label{fig_drivcav_stst_Re1200}
\end{figure}

\subsubsection{Transient Case with \lstinline$Re=800$ and Distributed Actuation}\label{ssec_drivcav_tdp_Re800_dcont}
\fbox{
\begin{minipage}[t]{.9\textwidth}
\begin{minipage}[t]{.6\textwidth}
	\begin{tabular}{l}
	  Setup: \hfill \lstinline$drivencavity$ \\
   Transient simulation with IMEX-Euler\\
   Reynolds number: \hfill \lstinline$Re=800$\\
   Grid: \hfill \lstinline$N=20$\\
   Time grid: \hfill \lstinline$t0, tE, Nts = 0., 20., 2**12$\\
	 Initial value: \hfill \emph{Stokes solution} \\
   Input frequency:\hfill \lstinline$omega=4*Pi/(tE-t0)$ \\ 
   Input: \hfill \lstinline$uvec=[sin(omega*t), cos(omega*t)].T$
 \end{tabular}
\end{minipage}
\hfill
\begin{minipage}[t]{.3\textwidth}
	\begin{tabular}{l}
  Simulation results: \\
  \lstinline$v(tk)$:\hfill Figure \ref{fig_drivcav_tdp_Re800}\\
  \lstinline$C*v$: \hfill Figure \ref{fig_drivcav_tdp_Re800_dcont_voutpout}\\
  \lstinline$C*p$: \hfill Figure \ref{fig_drivcav_tdp_Re800_dcont_voutpout}
 \end{tabular}
\end{minipage}
\\[.15in]
To replicate this simulation run
$$
  \mlsti{drivencavity_tdp_pout_vout_distcontrol.\{m,py\}}
$$
\end{minipage}
}

\begin{figure}[p]
	{\includegraphics[scale=0.22]{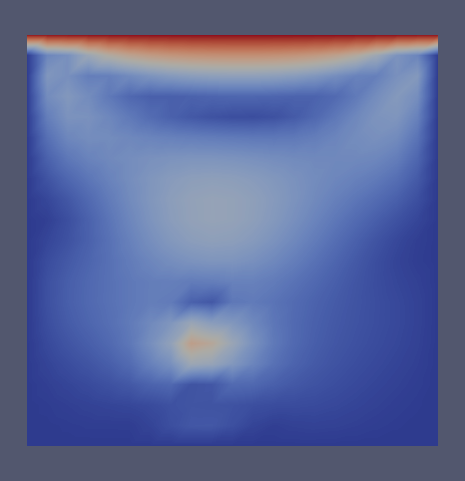}}
	{\includegraphics[scale=0.22]{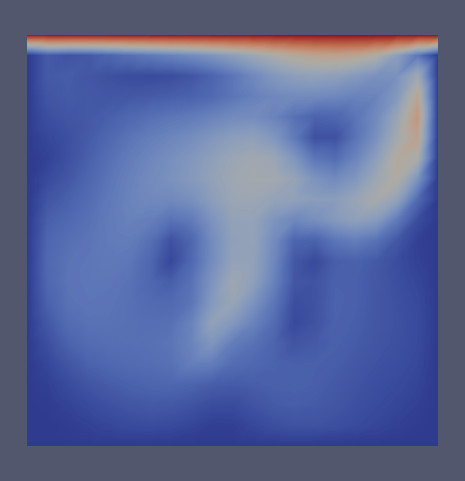}}
	{\includegraphics[scale=0.22]{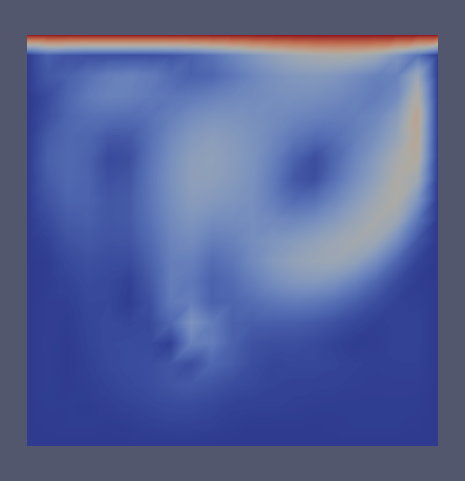}}\\[.04in]
	{\includegraphics[scale=0.22]{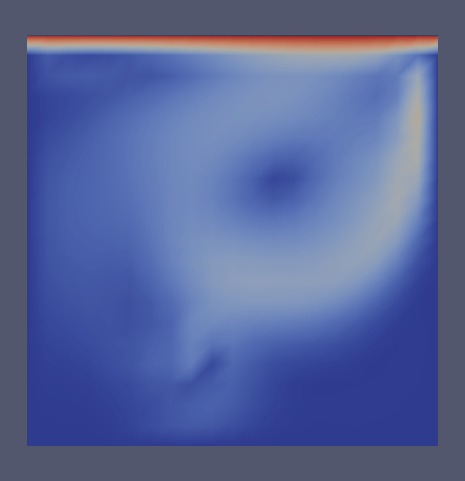}}
	{\includegraphics[scale=0.22]{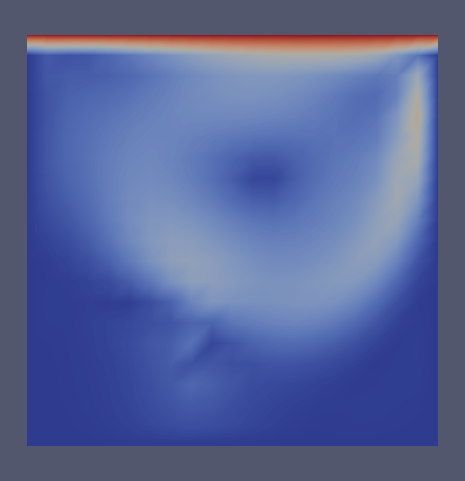}}
	{\includegraphics[scale=0.22]{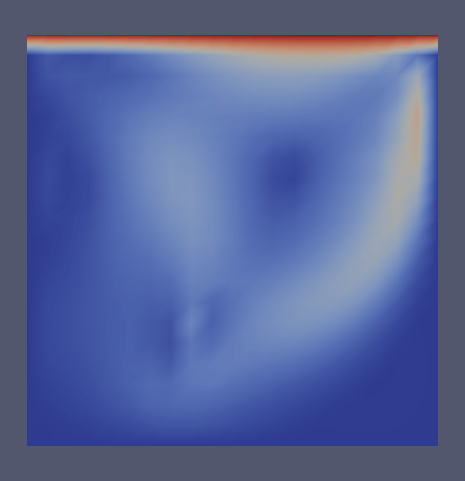}}\\[.04in]
	{\includegraphics[scale=0.22]{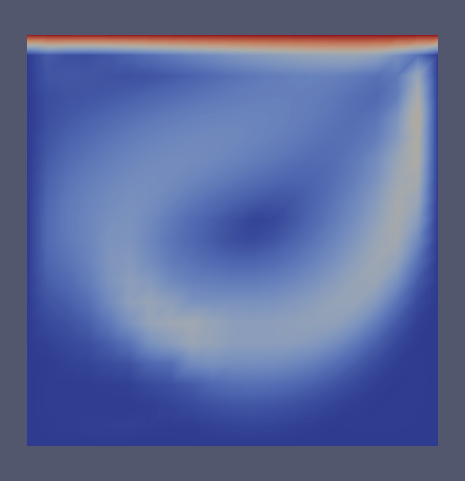}}
	{\includegraphics[scale=0.22]{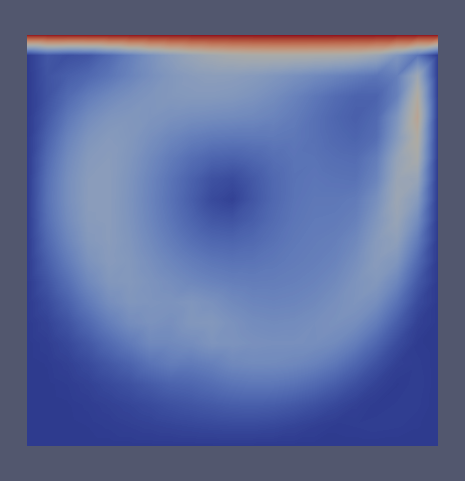}}
	{\includegraphics[scale=0.22]{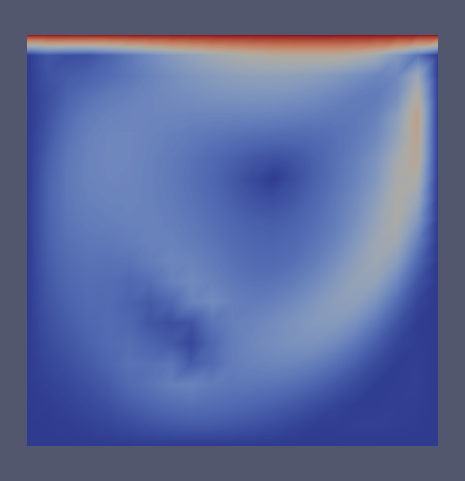}}
	\caption{Illustration of the velocity magnitude field for the setup described in Section \ref{ssec_drivcav_tdp_Re800_dcont} at \lstinline$t=0.0, 0.5, 1.0, 1.5, 2.0, 2.5, 3.0, 3.5, 4.0$.}
  \label{fig_drivcav_tdp_Re800}
\end{figure}

\setlength\figurewidth{6cm}
\setlength\figureheight{5cm}
\begin{figure}[p]
\input{plots/p_drivcav-N20-tE20.0-Nts4096-bccomg4.0.tikz}
\input{plots/v_drivcav-N20-tE20.0-Nts4096-bccomg4.0.tikz}
\caption{Illustration of the velocity and pressure output for the setup described in Section \ref{ssec_drivcav_tdp_Re800_dcont}}
  \label{fig_drivcav_tdp_Re800_dcont_voutpout}
\end{figure}

%% file: plots/p_nsequadtens-N3-tE4.0-Nts2048.tikz
% This file was created by matplotlib2tikz v0.5.3.
% The lastest updates can be retrieved from
% 
% https://github.com/nschloe/matplotlib2tikz
% 
% where you can also submit bug reports and leavecomments.
% 
\begin{tikzpicture}

\begin{axis}[
xmin=0, xmax=4,
ymin=0, ymax=0.00025,
axis on top,
width=\figurewidth,
height=\figureheight
]
\addplot [thick, red]
table {%
0.001953125 0.000166967849931637
0.01171875 0.000167117683356447
0.021484375 0.000167273521978749
0.03125 0.000167447779240911
0.041015625 0.000167644797951713
0.05078125 0.000167868566667143
0.060546875 0.000168122385569913
0.0703125 0.000168407766763761
0.080078125 0.000168727089653544
0.08984375 0.000169083101479728
0.099609375 0.000169478503974563
0.109375 0.000169915960403525
0.119140625 0.000170398348559788
0.12890625 0.000170929086763461
0.138671875 0.00017151205938985
0.1484375 0.000172151516144
0.158203125 0.000172852166775841
0.16796875 0.000173619291519727
0.177734375 0.000174458479158047
0.1875 0.000175375638052156
0.197265625 0.000176376811342808
0.20703125 0.000177467935044798
0.216796875 0.000178654794982615
0.2265625 0.000179942497028781
0.236328125 0.000181335403169109
0.24609375 0.000182836721948621
0.255859375 0.000184448361653949
0.265625 0.000186170473960853
0.275390625 0.000188000940262611
0.28515625 0.000189935212022117
0.294921875 0.000191966047834325
0.3046875 0.000194083423058314
0.314453125 0.000196275056194284
0.32421875 0.000198527123484817
0.333984375 0.000200825158046449
0.34375 0.000203154382689051
0.353515625 0.000205499801187276
0.36328125 0.0002078469861741
0.373046875 0.000210181012065231
0.3828125 0.000212483931550853
0.392578125 0.000214732728342749
0.40234375 0.000216896974213674
0.412109375 0.000218937017612752
0.421875 0.000220805402116141
0.431640625 0.000222450179170181
0.44140625 0.000223820530667217
0.451171875 0.000224875237781727
0.4609375 0.000225589669068877
0.470703125 0.000225960324438149
0.48046875 0.000226007712517341
0.490234375 0.000225773437841183
0.5 0.000225312313430626
0.509765625 0.000224684934125246
0.51953125 0.000223951935718267
0.529296875 0.00022317108224837
0.5390625 0.000222397248201172
0.548828125 0.000221681897643868
0.55859375 0.000221068729859149
0.568359375 0.000220586513100629
0.578125 0.000220243331260614
0.587890625 0.000220022200604701
0.59765625 0.000219881818129341
0.607421875 0.000219763688497534
0.6171875 0.000219600061471077
0.626953125 0.000219320992349812
0.63671875 0.000218861811370724
0.646484375 0.000218168335888644
0.65625 0.000217197988138693
0.666015625 0.00021592006047569
0.67578125 0.000214315957582287
0.685546875 0.000212379505550205
0.6953125 0.000210118161694679
0.705078125 0.000207554348518506
0.71484375 0.000204726564707581
0.724609375 0.000201690136423579
0.734375 0.000198517402332118
0.744140625 0.000195297103896178
0.75390625 0.00019213255280306
0.763671875 0.000189138433787846
0.7734375 0.000186436252618191
0.783203125 0.00018414849899015
0.79296875 0.000182391251100727
0.802734375 0.000181264998858294
0.8125 0.000180844746657477
0.822265625 0.000181170904846573
0.83203125 0.000182241868552721
0.841796875 0.000184009008978143
0.8515625 0.000186375310948285
0.861328125 0.000189198914512797
0.87109375 0.000192301612081117
0.880859375 0.000195481397798039
0.890625 0.000198527708540454
0.900390625 0.000201237399614126
0.91015625 0.000203429110703537
0.919921875 0.000204954127424348
0.9296875 0.000205702822663032
0.939453125 0.000205606678225344
0.94921875 0.000204636829809977
0.958984375 0.000202800712526703
0.96875 0.000200138335465611
0.978515625 0.000196719253395611
0.98828125 0.000192640675863883
0.998046875 0.000188026522380298
1.0078125 0.000183026746163963
1.017578125 0.000177815959245067
1.02734375 0.000172590377587721
1.037109375 0.00016756230833734
1.046875 0.000162951692038409
1.056640625 0.000158974584431804
1.06640625 0.000155828935634515
1.076171875 0.000153678558164912
1.0859375 0.000152636728197689
1.095703125 0.00015275137791901
1.10546875 0.000153994185204316
1.115234375 0.000156255901776451
1.125 0.000159349801056986
1.134765625 0.000163024065619383
1.14453125 0.000166982417978883
1.154296875 0.0001709106497278
1.1640625 0.000174505283595613
1.173828125 0.000177499778008159
1.18359375 0.000179683877020126
1.193359375 0.000180913240556575
1.203125 0.000181109064663115
1.212890625 0.000180249992511823
1.22265625 0.000178360233199724
1.232421875 0.00017549811422586
1.2421875 0.000171748363745875
1.251953125 0.000167219471311743
1.26171875 0.00016204523421327
1.271484375 0.000156388190412147
1.28125 0.000150442569243727
1.291015625 0.000144435049286139
1.30078125 0.000138622174823965
1.310546875 0.000133283570070751
1.3203125 0.000128710381218001
1.330078125 0.000125188859936711
1.33984375 0.000122979632266027
1.349609375 0.00012229388529998
1.359375 0.000123268479425364
1.369140625 0.000125942877135563
1.37890625 0.000130241577315337
1.388671875 0.000135966115892239
1.3984375 0.000142800278740822
1.408203125 0.00015033059967509
1.41796875 0.000158081323272719
1.427734375 0.000165559374779082
1.4375 0.000172301942307531
1.447265625 0.000177918272224194
1.45703125 0.000182118272673781
1.466796875 0.000184723419890447
1.4765625 0.000185660255492205
1.486328125 0.000184941599100048
1.49609375 0.000182642876996402
1.505859375 0.000178880515624505
1.515625 0.00017379748006494
1.525390625 0.000167558303564592
1.53515625 0.000160352638562996
1.544921875 0.000152403400048202
1.5546875 0.000143974437819894
1.564453125 0.000135373912818962
1.57421875 0.000126951690060266
1.583984375 0.000119090286869934
1.59375 0.000112188919866707
1.603515625 0.000106640002286696
1.61328125 0.000102798616296953
1.623046875 0.000100948019138547
1.6328125 0.000101265630565611
1.642578125 0.000103793429081704
1.65234375 0.000108416932810178
1.662109375 0.000114858422114634
1.671875 0.000122689713360915
1.681640625 0.000131366374310527
1.69140625 0.000140280236527431
1.701171875 0.000148822513314212
1.7109375 0.000156447589391227
1.720703125 0.000162726391268974
1.73046875 0.000167377264082285
1.740234375 0.000170266543798961
1.75 0.000171382500156009
1.759765625 0.000170795999169092
1.76953125 0.000168621714257263
1.779296875 0.000164989821242356
1.7890625 0.000160035140050401
1.798828125 0.000153904848603185
1.80859375 0.000146776959511973
1.818359375 0.000138878389096415
1.828125 0.000130495680448565
1.837890625 0.000121976497792118
1.84765625 0.00011372280599392
1.857421875 0.000106177422394324
1.8671875 9.98040876942465e-05
1.876953125 9.50597198327497e-05
1.88671875 9.23584973868289e-05
1.896484375 9.20293873943859e-05
1.90625 9.42708983136464e-05
1.916015625 9.91100873226064e-05
1.92578125 0.000106374778586356
1.935546875 0.000115686359031315
1.9453125 0.00012647811984411
1.955078125 0.000138042314812148
1.96484375 0.000149604222114859
1.974609375 0.000160411843653156
1.984375 0.000169821370401896
1.994140625 0.000177358303236504
2.00390625 0.000182741545468271
2.013671875 0.000185868971530348
2.0234375 0.000186774704797843
2.033203125 0.000185574507103378
2.04296875 0.000182415736458856
2.052734375 0.000177446761589574
2.0625 0.00017081395786245
2.072265625 0.0001626808090819
2.08203125 0.000153253842529559
2.091796875 0.000142803317747053
2.1015625 0.000131675822128906
2.111328125 0.00012029817614199
2.12109375 0.000109170249431371
2.130859375 9.88462237797837e-05
2.140625 8.99061907907345e-05
2.150390625 8.29190376982423e-05
2.16015625 7.83952892450053e-05
2.169921875 7.67308575403208e-05
2.1796875 7.81504133843944e-05
2.189453125 8.26621371757746e-05
2.19921875 9.00314738816597e-05
2.208984375 9.97800655330719e-05
2.21875 0.000111217112226641
2.228515625 0.000123505579770981
2.23828125 0.000135756351136082
2.248046875 0.00014713597061377
2.2578125 0.000156967066649294
2.267578125 0.000164794264657085
2.27734375 0.000170394691889107
2.287109375 0.000173735690031259
2.296875 0.000174903394425833
2.306640625 0.000174031053772423
2.31640625 0.000171251433290767
2.326171875 0.00016668479134866
2.3359375 0.000160454445234632
2.345703125 0.000152713303431222
2.35546875 0.000143671599919321
2.365234375 0.00013362043112038
2.375 0.000122944041844453
2.384765625 0.000112117368099336
2.39453125 0.000101691639316318
2.404296875 9.22725424957056e-05
2.4140625 8.44913391708529e-05
2.423828125 7.8962454872814e-05
2.43359375 7.62233566687582e-05
2.443359375 7.6665899384082e-05
2.453125 8.04742354500841e-05
2.462890625 8.75772916986969e-05
2.47265625 9.76207685754425e-05
2.482421875 0.000109969883721995
2.4921875 0.000123757237172129
2.501953125 0.000137981086403732
2.51171875 0.000151636903471938
2.521484375 0.000163845710480527
2.53125 0.000173948211821328
2.541015625 0.000181550180241804
2.55078125 0.000186510818994313
2.560546875 0.000188877132676031
2.5703125 0.000188790192694536
2.580078125 0.000186402317545143
2.58984375 0.000181836356616955
2.599609375 0.000175195096928338
2.609375 0.000166601267179122
2.619140625 0.000156235760477115
2.62890625 0.000144359772095037
2.638671875 0.000131330578149836
2.6484375 0.000117615553583146
2.658203125 0.000103794025677253
2.66796875 9.0539425293039e-05
2.677734375 7.85856489599973e-05
2.6875 6.86835136614748e-05
2.697265625 6.15441069548073e-05
2.70703125 5.77650412286295e-05
2.716796875 5.77510641093636e-05
2.7265625 6.16488112717088e-05
2.736328125 6.93050344556409e-05
2.74609375 8.02502036667569e-05
2.755859375 9.3718078695195e-05
2.765625 0.000108714295005451
2.775390625 0.00012413080975312
2.78515625 0.000138888512496525
2.794921875 0.00015207962725283
2.8046875 0.000163068774344867
2.814453125 0.000171520784184671
2.82421875 0.00017735437849798
2.833984375 0.000180645504221958
2.84375 0.000181522600926629
2.853515625 0.000180101547404544
2.86328125 0.000176477430709926
2.873046875 0.000170748870666816
2.8828125 0.000163048105044524
2.892578125 0.000153573046949094
2.90234375 0.000142619006084246
2.912109375 0.000130597364556359
2.921875 0.000118035643226062
2.931640625 0.000105565666948395
2.94140625 9.39052671438032e-05
2.951171875 8.38274538261329e-05
2.9609375 7.61064096335929e-05
2.970703125 7.1441670158536e-05
2.98046875 7.03752634587113e-05
2.990234375 7.32154438052628e-05
3 7.9973756145959e-05
3.009765625 9.03245429128838e-05
3.01953125 0.00010360141346775
3.029296875 0.000118845501254987
3.0390625 0.000134916902945387
3.048828125 0.00015065592980171
3.05859375 0.000165041883094191
3.068359375 0.000177302302611564
3.078125 0.000186963269924296
3.087890625 0.000193836031759565
3.09765625 0.000197935825921519
3.107421875 0.000199365762821203
3.1171875 0.000198223779938099
3.126953125 0.000194569030028883
3.13671875 0.000188446473028879
3.146484375 0.00017993853002791
3.15625 0.000169203397501485
3.166015625 0.000156492438192267
3.17578125 0.000142169239113919
3.185546875 0.000126733295688841
3.1953125 0.000110827765077443
3.205078125 9.52197874155327e-05
3.21484375 8.0761384256833e-05
3.224609375 6.83421965404108e-05
3.234375 5.88272454835657e-05
3.244140625 5.29662679122494e-05
3.25390625 5.12888854408029e-05
3.263671875 5.40197317091691e-05
3.2734375 6.1028056439884e-05
3.283203125 7.18090518206823e-05
3.29296875 8.55111971805787e-05
3.302734375 0.000101026764169825
3.3125 0.000117132951316958
3.322265625 0.000132659362140514
3.33203125 0.000146649164076727
3.341796875 0.000158461449249044
3.3515625 0.000167783661541861
3.361328125 0.000174563129324797
3.37109375 0.000178884376301744
3.380859375 0.000180845287338612
3.390625 0.000180496880426953
3.400390625 0.000177858853692512
3.41015625 0.000172963137468801
3.419921875 0.000165888038735282
3.4296875 0.000156788956078709
3.439453125 0.000145933046440817
3.44921875 0.000133723581707579
3.458984375 0.000120703368242907
3.46875 0.000107544585859433
3.478515625 9.50325695624573e-05
3.48828125 8.40319182876726e-05
3.498046875 7.54201343074981e-05
3.5078125 6.99974269546856e-05
3.517578125 6.839178391583e-05
3.52734375 7.09670075062165e-05
3.537109375 7.77436528257374e-05
3.546875 8.83550451124487e-05
3.556640625 0.00010205188130734
3.56640625 0.000117762484564146
3.576171875 0.000134227939304226
3.5859375 0.000150201006880336
3.595703125 0.000164632137560122
3.60546875 0.000176777981129737
3.615234375 0.000186232172150292
3.625 0.000192883629301285
3.634765625 0.000196804130757351
3.64453125 0.000198112482624693
3.654296875 0.000196884296457199
3.6640625 0.000193137886344011
3.673828125 0.000186879424764277
3.68359375 0.000178162064601956
3.693359375 0.000167119557039246
3.703125 0.000153983250963502
3.712890625 0.000139110781466978
3.72265625 0.000123017488803782
3.732421875 0.000106383428207257
3.7421875 9.00298019247622e-05
3.751953125 7.48767722359203e-05
3.76171875 6.18926543549039e-05
3.771484375 5.2023406789887e-05
3.78125 4.60862032912827e-05
3.791015625 4.46478419358836e-05
3.80078125 4.79305697220595e-05
3.810546875 5.57562353196205e-05
3.8203125 6.75243099583172e-05
3.830078125 8.22501493849522e-05
3.83984375 9.86810346119454e-05
3.849609375 0.000115466947768607
3.859375 0.000131357893097761
3.869140625 0.000145381466496507
3.87890625 0.000156936596879644
3.888671875 0.000165784388550099
3.8984375 0.000171954452436321
3.908203125 0.00017559449551408
3.91796875 0.000176831923344549
3.927734375 0.000175724598980992
3.9375 0.000172297471341332
3.947265625 0.000166595073106591
3.95703125 0.000158712878490092
3.966796875 0.000148826231166501
3.9765625 0.000137226440127555
3.986328125 0.000124346473000918
3.99609375 0.000110766401114852
};
\end{axis}

\end{tikzpicture}

%% file: plots/p_nsequadtns-N2-tE6.0-Nts1024-bccomg3.0.tikz
% This file was created by matplotlib2tikz v0.5.3.
% The lastest updates can be retrieved from
% 
% https://github.com/nschloe/matplotlib2tikz
% 
% where you can also submit bug reports and leavecomments.
% 
\begin{tikzpicture}

\begin{axis}[
xmin=0, xmax=6,
ymin=0.0012, ymax=0.0028,
axis on top,
width=\figurewidth,
height=\figureheight
]
\addplot [thick, red]
table {%
0.005859375 0.00269172420051011
0.03515625 0.00261889708845293
0.064453125 0.00242449412719735
0.09375 0.00223877214690521
0.123046875 0.00208330614272081
0.15234375 0.00195728815116115
0.181640625 0.0018504999239058
0.2109375 0.00175413256478432
0.240234375 0.00167820509345082
0.26953125 0.00163081763453703
0.298828125 0.00160064899852568
0.328125 0.00158287078870839
0.357421875 0.00157449517112902
0.38671875 0.00157235534197537
0.416015625 0.00157361892438196
0.4453125 0.00157486571916963
0.474609375 0.00157271905532301
0.50390625 0.00156549442913781
0.533203125 0.00155426553323785
0.5625 0.00154212156476195
0.591796875 0.00153310375376237
0.62109375 0.00153041759687855
0.650390625 0.00153352170742163
0.6796875 0.00153890055189191
0.708984375 0.0015441622367649
0.73828125 0.00154902771173472
0.767578125 0.00155304762877632
0.796875 0.0015561749605722
0.826171875 0.00155803120596043
0.85546875 0.00155809055955509
0.884765625 0.00155560134338152
0.9140625 0.00155007400051683
0.943359375 0.00154146566918
0.97265625 0.00153032478085843
1.001953125 0.00151765105181708
1.03125 0.00150465608169595
1.060546875 0.00149241642322164
1.08984375 0.00148163108721216
1.119140625 0.00147249952117766
1.1484375 0.00146479305558179
1.177734375 0.00145802797734274
1.20703125 0.00145168895582291
1.236328125 0.00144540220557418
1.265625 0.00143900876780859
1.294921875 0.00143254148202511
1.32421875 0.00142614151268484
1.353515625 0.00141996888868132
1.3828125 0.0014141393820183
1.412109375 0.00140870189999398
1.44140625 0.00140366205625101
1.470703125 0.00139903280611224
1.5 0.00139488312990869
1.529296875 0.00139136065493504
1.55859375 0.00138867534169337
1.587890625 0.00138705159927809
1.6171875 0.00138667139865162
1.646484375 0.00138763383045901
1.67578125 0.00138993997048808
1.705078125 0.00139350397520447
1.734375 0.00139818806185607
1.763671875 0.00140384994743642
1.79296875 0.00141037997495941
1.822265625 0.00141771415924786
1.8515625 0.00142582745452882
1.880859375 0.00143471620596809
1.91015625 0.00144437394011301
1.939453125 0.00145476787771993
1.96875 0.0014658279228054
1.998046875 0.00147745190020207
2.02734375 0.0014895211855913
2.056640625 0.00150191949849909
2.0859375 0.00151454850159644
2.115234375 0.0015273312896195
2.14453125 0.00154020411215468
2.173828125 0.00155310682879324
2.203125 0.00156597431849054
2.232421875 0.00157872873908118
2.26171875 0.00159127448025174
2.291015625 0.00160349965219659
2.3203125 0.0016152917492661
2.349609375 0.00162655410294094
2.37890625 0.00163721297456497
2.408203125 0.00164721598620508
2.4375 0.00165652682650844
2.466796875 0.00166511547114855
2.49609375 0.00167294765263689
2.525390625 0.00167998058297006
2.5546875 0.001686163577841
2.583984375 0.00169144128762591
2.61328125 0.00169575959084217
2.642578125 0.00169907179402087
2.671875 0.00170134206290453
2.701171875 0.00170254555499885
2.73046875 0.0017026661551482
2.759765625 0.00170169340052207
2.7890625 0.00169962032544885
2.818359375 0.00169644320786543
2.84765625 0.00169216356917788
2.876953125 0.00168679206785366
2.90625 0.00168035294242033
2.935546875 0.00167288731560968
2.96484375 0.00166445419118021
2.994140625 0.00165512897798391
3.0234375 0.00164500033125339
3.052734375 0.00163416654039674
3.08203125 0.00162273255665278
3.111328125 0.00161080825608949
3.140625 0.00159850793599797
3.169921875 0.00158595054713985
3.19921875 0.00157325990469173
3.228515625 0.0015605641753112
3.2578125 0.00154799426248109
3.287109375 0.00153568114397786
3.31640625 0.00152375257188652
3.345703125 0.00151232971081482
3.375 0.0015015242473002
3.404296875 0.00149143631145583
3.43359375 0.0014821532993392
3.462890625 0.00147374945968551
3.4921875 0.00146628598373862
3.521484375 0.00145981134227052
3.55078125 0.00145436171734384
3.580078125 0.00144996150547472
3.609375 0.00144662395368914
3.638671875 0.00144435199990356
3.66796875 0.00144313933357587
3.697265625 0.00144297160667764
3.7265625 0.00144382764844824
3.755859375 0.00144568050432787
3.78515625 0.00144849814638857
3.814453125 0.0014522437835139
3.84375 0.00145687580743569
3.873046875 0.00146234750790915
3.90234375 0.00146860674510472
3.931640625 0.00147559576551185
3.9609375 0.00148325129604063
3.990234375 0.00149150497104355
4.01953125 0.00150028406598592
4.048828125 0.00150951245197191
4.078125 0.00151911165991407
4.107421875 0.00152900194876676
4.13671875 0.00153910329848576
4.166015625 0.00154933628022549
4.1953125 0.00155962278267766
4.224609375 0.00156988658689393
4.25390625 0.00158005378441168
4.283203125 0.00159005303170071
4.3125 0.00159981563595381
4.341796875 0.00160927547637642
4.37109375 0.00161836878219512
4.400390625 0.00162703381068291
4.4296875 0.00163521048793167
4.458984375 0.00164284008474488
4.48828125 0.00164986499649959
4.517578125 0.0016562286805781
4.546875 0.00166187578248535
4.576171875 0.0016667524575527
4.60546875 0.0016708068762106
4.634765625 0.00167398988978484
4.6640625 0.00167625583102056
4.693359375 0.00167756342576451
4.72265625 0.00167787679633301
4.751953125 0.00167716653863122
4.78125 0.00167541085249174
4.810546875 0.00167259669749175
4.83984375 0.00166872093712034
4.869140625 0.00166379142446197
4.8984375 0.00165782797527716
4.927734375 0.00165086317064508
4.95703125 0.00164294293159298
4.986328125 0.0016341268117473
5.015625 0.00162448796063622
5.044921875 0.00161411271955734
5.07421875 0.00160309982447793
5.103515625 0.00159155920670174
5.1328125 0.00157961040241772
5.162109375 0.00156738060591132
5.19140625 0.00155500242638131
5.220703125 0.00154261143182453
5.25 0.0015303435820222
5.279296875 0.00151833266306776
5.30859375 0.00150670783621579
5.337890625 0.00149559140355911
5.3671875 0.00148509687387366
5.396484375 0.00147532738634393
5.42578125 0.00146637452171482
5.455078125 0.00145831750298389
5.484375 0.00145122276433319
5.513671875 0.0014451438492091
5.54296875 0.00144012158717473
5.572265625 0.00143618449414273
5.6015625 0.0014333493407655
5.630859375 0.00143162183785389
5.66015625 0.00143099739429956
5.689453125 0.00143146191081063
5.71875 0.00143299258102032
5.748046875 0.00143555867927087
5.77734375 0.0014391223212228
5.806640625 0.00144363918944555
5.8359375 0.00144905922043544
5.865234375 0.00145532725305717
5.89453125 0.00146238364036251
5.923828125 0.00147016482798697
5.953125 0.00147860390269262
5.982421875 0.00148763111433831
};
\end{axis}

\end{tikzpicture}

%% file: plots/p_drivcav-N20-tE20.0-Nts4096-bccomg4.0.tikz
% This file was created by matplotlib2tikz v0.5.3.
% The lastest updates can be retrieved from
% 
% https://github.com/nschloe/matplotlib2tikz
% 
% where you can also submit bug reports and leavecomments.
% 
\begin{tikzpicture}

\begin{axis}[
xmin=0, xmax=20,
ymin=-0.0007, ymax=-9.99999999999999e-05,
axis on top,
width=\figurewidth,
height=\figureheight
]
\addplot [thick, red]
table {%
0.0048828125 -0.000621366992389437
0.029296875 -0.000617654819848506
0.0537109375 -0.000614545894829579
0.078125 -0.00061177806531266
0.1025390625 -0.000609137350504432
0.126953125 -0.000606511288190391
0.1513671875 -0.000603851887602539
0.17578125 -0.00060111535833451
0.2001953125 -0.000598232445024957
0.224609375 -0.0005951201978492
0.2490234375 -0.000591709734662853
0.2734375 -0.000587960189881089
0.2978515625 -0.000583851481594188
0.322265625 -0.000579370767523297
0.3466796875 -0.000574507955966682
0.37109375 -0.000569259197300336
0.3955078125 -0.000563626213277804
0.419921875 -0.000557606706239119
0.4443359375 -0.000551185928732166
0.46875 -0.000544341566623076
0.4931640625 -0.000537060585058414
0.517578125 -0.00052935414543506
0.5419921875 -0.000521258644046891
0.56640625 -0.000512823630990698
0.5908203125 -0.000504097378107371
0.615234375 -0.000495120114173889
0.6396484375 -0.000485926671719245
0.6640625 -0.00047655303302656
0.6884765625 -0.000467040213206938
0.712890625 -0.000457432896762672
0.7373046875 -0.000447774757318438
0.76171875 -0.000438104131138582
0.7861328125 -0.000428452481724617
0.810546875 -0.000418845717662628
0.8349609375 -0.000409306769407481
0.859375 -0.000399857621133389
0.8837890625 -0.000390519847264911
0.908203125 -0.000381313779087534
0.9326171875 -0.000372257074973005
0.95703125 -0.000363363511047054
0.9814453125 -0.000354642451602866
1.005859375 -0.000346099027305778
1.0302734375 -0.000337734778371021
1.0546875 -0.000329548471804464
1.0791015625 -0.000321536904134448
1.103515625 -0.000313695634522373
1.1279296875 -0.000306019669246232
1.15234375 -0.000298504117251612
1.1767578125 -0.00029114479566161
1.201171875 -0.000283938736908714
1.2255859375 -0.000276884561020531
1.25 -0.000269982713345096
1.2744140625 -0.000263235596958135
1.298828125 -0.000256647629697041
1.3232421875 -0.0002502252336312
1.34765625 -0.000243976741360221
1.3720703125 -0.000237912197225103
1.396484375 -0.000232043045363037
1.4208984375 -0.000226381720382464
1.4453125 -0.000220941176393883
1.4697265625 -0.000215734397807668
1.494140625 -0.000210773930375587
1.5185546875 -0.00020607145876429
1.54296875 -0.000201637443956231
1.5673828125 -0.000197480824240509
1.591796875 -0.000193608778900578
1.6162109375 -0.000190026552745654
1.640625 -0.000186737340461394
1.6650390625 -0.000183742230444195
1.689453125 -0.000181040207368893
1.7138671875 -0.000178628211151528
1.73828125 -0.000176501247739463
1.7626953125 -0.000174652544995977
1.787109375 -0.000173073745440043
1.8115234375 -0.000171755126809882
1.8359375 -0.000170685841419229
1.8603515625 -0.000169854165646958
1.884765625 -0.000169247751529987
1.9091796875 -0.000168853872897049
1.93359375 -0.00016865965886542
1.9580078125 -0.000168652307594426
1.982421875 -0.000168819273302535
2.0068359375 -0.000169148419770296
2.03125 -0.000169628134291925
2.0556640625 -0.000170247397413442
2.080078125 -0.000170995806012124
2.1044921875 -0.00017186355015727
2.12890625 -0.00017284134757785
2.1533203125 -0.000173920342917623
2.177734375 -0.000175091981961953
2.2021484375 -0.000176347873069176
2.2265625 -0.000177679648961068
2.2509765625 -0.000179078841564713
2.275390625 -0.00018053678086398
2.2998046875 -0.000182044525962178
2.32421875 -0.000183592833115987
2.3486328125 -0.000185172161939425
2.373046875 -0.000186772717671744
2.3974609375 -0.000188384524771724
2.421875 -0.000189997525380824
2.4462890625 -0.000191601695367884
2.470703125 -0.000193187170750811
2.4951171875 -0.000194744377910392
2.51953125 -0.000196264162102318
2.5439453125 -0.000197737909956422
2.568359375 -0.000199157662861012
2.5927734375 -0.000200516219170141
2.6171875 -0.000201807224057701
2.6416015625 -0.000203025246443205
2.666015625 -0.000204165842815772
2.6904296875 -0.000205225607903556
2.71484375 -0.000206202212075408
2.7392578125 -0.000207094425133862
2.763671875 -0.000207902125911056
2.7880859375 -0.000208626296862487
2.8125 -0.000209269002800987
2.8369140625 -0.000209833353081996
2.861328125 -0.000210323446870753
2.8857421875 -0.000210744301711871
2.91015625 -0.000211101766192746
2.9345703125 -0.000211402418192436
2.958984375 -0.000211653450759778
2.9833984375 -0.000211862548149503
3.0078125 -0.000212037754888151
3.0322265625 -0.000212187340938054
3.056640625 -0.000212319666074329
3.0810546875 -0.000212443046575853
3.10546875 -0.000212565627148127
3.1298828125 -0.000212695260771868
3.154296875 -0.000212839398858685
3.1787109375 -0.000213004993698607
3.203125 -0.000213198414754969
3.2275390625 -0.000213425379894435
3.251953125 -0.000213690902144512
3.2763671875 -0.000213999252118107
3.30078125 -0.000214353935804353
3.3251953125 -0.000214757687040721
3.349609375 -0.00021521247367518
3.3740234375 -0.00021571951619371
3.3984375 -0.000216279317404816
3.4228515625 -0.000216891701707765
3.447265625 -0.000217555862401632
3.4716796875 -0.000218270415542421
3.49609375 -0.000219033458877408
3.5205078125 -0.000219842634502544
3.544921875 -0.000220695193946535
3.5693359375 -0.000221588064526921
3.59375 -0.000222517915909545
3.6181640625 -0.000223481225914172
3.642578125 -0.000224474344709577
3.6669921875 -0.000225493556627209
3.69140625 -0.000226535138920255
3.7158203125 -0.000227595416862554
3.740234375 -0.000228670814671991
3.7646484375 -0.000229757901817752
3.7890625 -0.00023085343434524
3.8134765625 -0.000231954390935339
3.837890625 -0.000233058003498075
3.8623046875 -0.000234161782182946
3.88671875 -0.000235263534772355
3.9111328125 -0.000236361380507297
3.935546875 -0.000237453758481952
3.9599609375 -0.00023853943080973
3.984375 -0.000239617480842065
4.0087890625 -0.000240687306781886
4.033203125 -0.00024174861107398
4.0576171875 -0.000242801386018638
4.08203125 -0.00024384589605703
4.1064453125 -0.000244882657227139
4.130859375 -0.00024591241427994
4.1552734375 -0.000246936115969896
4.1796875 -0.000247954889009515
4.2041015625 -0.000248970011198889
4.228515625 -0.000249982884210338
4.2529296875 -0.000250995006502969
4.27734375 -0.000252007946815584
4.3017578125 -0.000253023318677968
4.326171875 -0.000254042756324575
4.3505859375 -0.000255067892382473
4.375 -0.000256100337642997
4.3994140625 -0.000257141663175491
4.423828125 -0.000258193384975268
4.4482421875 -0.000259256951270356
4.47265625 -0.000260333732522335
4.4970703125 -0.000261425014080252
4.521484375 -0.000262531991358388
4.5458984375 -0.000263655767329118
4.5703125 -0.000264797352052954
4.5947265625 -0.000265957663902121
4.619140625 -0.000267137532097032
4.6435546875 -0.000268337700154035
4.66796875 -0.000269558829829545
4.6923828125 -0.000270801505177791
4.716796875 -0.000272066236373553
4.7412109375 -0.000273353463001265
4.765625 -0.000274663556600784
4.7900390625 -0.000275996822311577
4.814453125 -0.000277353499562651
4.8388671875 -0.000278733761822362
4.86328125 -0.000280137715499589
4.8876953125 -0.000281565398147049
4.912109375 -0.000283016776163044
4.9365234375 -0.00028449174223875
4.9609375 -0.00028599011279074
4.9853515625 -0.00028751162564858
5.009765625 -0.000289055938242198
5.0341796875 -0.000290622626534059
5.05859375 -0.000292211184891892
5.0830078125 -0.000293821027087959
5.107421875 -0.000295451488558911
5.1318359375 -0.000297101830022554
5.15625 -0.000298771242489946
5.1806640625 -0.000300458853676567
5.205078125 -0.000302163735730636
5.2294921875 -0.00030388491416174
5.25390625 -0.000305621377773875
5.2783203125 -0.000307372089357841
5.302734375 -0.000309135996836656
5.3271484375 -0.000310912044534599
5.3515625 -0.000312699184201969
5.3759765625 -0.000314496385446583
5.400390625 -0.000316302645239943
5.4248046875 -0.000318116996208384
5.44921875 -0.000319938513508505
5.4736328125 -0.000321766320148844
5.498046875 -0.000323599590709397
5.5224609375 -0.000325437553516076
5.546875 -0.000327279491359817
5.5712890625 -0.000329124740944236
5.595703125 -0.000330972691243322
5.6201171875 -0.0003328227809751
5.64453125 -0.00033467449538422
5.6689453125 -0.000336527362496667
5.693359375 -0.000338380948992407
5.7177734375 -0.000340234855807997
5.7421875 -0.000342088713572682
5.7666015625 -0.000343942177974162
5.791015625 -0.000345794925170402
5.8154296875 -0.000347646647351875
5.83984375 -0.000349497048608671
5.8642578125 -0.000351345841252051
5.888671875 -0.000353192742725986
5.9130859375 -0.000355037473256686
5.9375 -0.000356879754316284
5.9619140625 -0.000358719307948327
5.986328125 -0.000360555856925707
6.0107421875 -0.000362389125648552
6.03515625 -0.000364218841625831
6.0595703125 -0.000366044737331216
6.083984375 -0.000367866552212359
6.1083984375 -0.000369684034609883
6.1328125 -0.000371496943380747
6.1572265625 -0.000373305049063667
6.181640625 -0.000375108134474126
6.2060546875 -0.000376905994705108
6.23046875 -0.000378698436554283
6.2548828125 -0.0003804852774921
6.279296875 -0.000382266344291924
6.3037109375 -0.000384041471512618
6.328125 -0.000385810499993473
6.3525390625 -0.0003875732755404
6.376953125 -0.000389329647936268
6.4013671875 -0.000391079470379357
6.42578125 -0.000392822599396543
6.4501953125 -0.000394558895253174
6.474609375 -0.000396288222816772
6.4990234375 -0.00039801045281142
6.5234375 -0.000399725463371998
6.5478515625 -0.000401433141777476
6.572265625 -0.000403133386275532
6.5966796875 -0.000404826107858085
6.62109375 -0.000406511231915256
6.6455078125 -0.000408188699677973
6.669921875 -0.000409858469380456
6.6943359375 -0.000411520517112411
6.71875 -0.000413174837337773
6.7431640625 -0.000414821443066672
6.767578125 -0.000416460365712249
6.7919921875 -0.000418091654631347
6.81640625 -0.000419715376409666
6.8408203125 -0.000421331613901432
6.865234375 -0.000422940465093179
6.8896484375 -0.000424542041807748
6.9140625 -0.000426136468306137
6.9384765625 -0.000427723879815771
6.962890625 -0.000429304421016192
6.9873046875 -0.000430878244521098
7.01171875 -0.000432445509365785
7.0361328125 -0.000434006379529233
7.060546875 -0.00043556102250261
7.0849609375 -0.000437109607914472
7.109375 -0.00043865230621747
7.1337890625 -0.000440189287445521
7.158203125 -0.00044172072004192
7.1826171875 -0.000443246769754678
7.20703125 -0.000444767598603796
7.2314453125 -0.000446283363918971
7.255859375 -0.000447794217435889
7.2802734375 -0.000449300304462799
7.3046875 -0.000450801763103688
7.3291015625 -0.000452298723540102
7.353515625 -0.00045379130736842
7.3779296875 -0.000455279626998526
7.40234375 -0.000456763785092824
7.4267578125 -0.000458243874070985
7.451171875 -0.000459719975654435
7.4755859375 -0.000461192160468003
7.5 -0.000462660487685525
7.5244140625 -0.000464125004730415
7.548828125 -0.000465585747011976
7.5732421875 -0.000467042737721873
7.59765625 -0.000468495987661692
7.6220703125 -0.000469945495118576
7.646484375 -0.000471391245785098
7.6708984375 -0.000472833212707679
7.6953125 -0.000474271356281276
7.7197265625 -0.0004757056242667
7.744140625 -0.000477135951844196
7.7685546875 -0.000478562261692716
7.79296875 -0.000479984464089675
7.8173828125 -0.000481402457033665
7.841796875 -0.00048281612638168
7.8662109375 -0.00048422534600063
7.890625 -0.000485629977933265
7.9150390625 -0.000487029872561639
7.939453125 -0.000488424868787281
7.9638671875 -0.000489814794204644
7.98828125 -0.000491199465283669
8.0126953125 -0.000492578687536221
8.037109375 -0.000493952255698349
8.0615234375 -0.000495319953893686
8.0859375 -0.000496681555796665
8.1103515625 -0.000498036824788328
8.134765625 -0.000499385514112283
8.1591796875 -0.00050072736701783
8.18359375 -0.000502062116899042
8.2080078125 -0.000503389487435393
8.232421875 -0.000504709192723198
8.2568359375 -0.000506020937407592
8.28125 -0.000507324416818628
8.3056640625 -0.000508619317102857
8.330078125 -0.000509905315365711
8.3544921875 -0.000511182079817561
8.37890625 -0.000512449269932459
8.4033203125 -0.000513706536617178
8.427734375 -0.000514953522395917
8.4521484375 -0.000516189861615497
8.4765625 -0.000517415180669861
8.5009765625 -0.000518629098247942
8.525390625 -0.00051983122561053
8.5498046875 -0.00052102116689708
8.57421875 -0.000522198519458034
8.5986328125 -0.000523362874228568
8.623046875 -0.000524513816131338
8.6474609375 -0.000525650924517001
8.671875 -0.00052677377364459
8.6962890625 -0.000527881933193848
8.720703125 -0.00052897496881797
8.7451171875 -0.000530052442731046
8.76953125 -0.000531113914329979
8.7939453125 -0.000532158940849402
8.818359375 -0.00053318707804458
8.8427734375 -0.00053419788090272
8.8671875 -0.000535190904373266
8.8916015625 -0.000536165704121875
8.916015625 -0.000537121837289357
8.9404296875 -0.000538058863264701
8.96484375 -0.00053897634444859
8.9892578125 -0.000539873847025337
9.013671875 -0.000540750941707751
9.0380859375 -0.000541607204464435
9.0625 -0.000542442217226405
9.0869140625 -0.000543255568541522
9.111328125 -0.000544046854196928
9.1357421875 -0.000544815677782472
9.16015625 -0.00054556165119678
9.1845703125 -0.000546284395080769
9.208984375 -0.00054698353918554
9.2333984375 -0.00054765872265351
9.2578125 -0.000548309594224658
9.2822265625 -0.000548935812353666
9.306640625 -0.000549537045243912
9.3310546875 -0.000550112970795937
9.35546875 -0.000550663276474979
9.3798828125 -0.00055118765910841
9.404296875 -0.000551685824603942
9.4287109375 -0.000552157487617584
9.453125 -0.000552602371168746
9.4775390625 -0.000553020206215363
9.501953125 -0.000553410731204271
9.5263671875 -0.000553773691616134
9.55078125 -0.000554108839499181
9.5751953125 -0.000554415933025835
9.599609375 -0.000554694736074456
9.6240234375 -0.0005549450178405
9.6484375 -0.000555166552503578
9.6728515625 -0.000555359118937426
9.697265625 -0.000555522500487477
9.7216796875 -0.00055565648480424
9.74609375 -0.000555760863741238
9.7705078125 -0.000555835433314944
9.794921875 -0.000555879993727988
9.8193359375 -0.000555894349439572
9.84375 -0.000555878309291479
9.8681640625 -0.0005558316866713
9.892578125 -0.000555754299707174
9.9169921875 -0.00055564597149404
9.94140625 -0.000555506530326221
9.9658203125 -0.000555335809937767
9.990234375 -0.000555133649742976
10.0146484375 -0.000554899895058192
10.0390625 -0.000554634397309205
10.0634765625 -0.000554337014206253
10.087890625 -0.000554007609888558
10.1123046875 -0.000553646055027967
10.13671875 -0.000553252226890859
10.1611328125 -0.000552826009356551
10.185546875 -0.000552367292885071
10.2099609375 -0.000551875974443083
10.234375 -0.000551351957380699
10.2587890625 -0.0005507951512638
10.283203125 -0.000550205471669188
10.3076171875 -0.000549582839939536
10.33203125 -0.000548927182904873
10.3564453125 -0.000548238432579228
10.380859375 -0.000547516525829669
10.4052734375 -0.000546761404031884
10.4296875 -0.00054597301270965
10.4541015625 -0.000545151301169279
10.478515625 -0.000544296222133036
10.5029296875 -0.000543407731369386
10.52734375 -0.000542485787341807
10.5517578125 -0.000541530350855927
10.576171875 -0.000540541384734171
10.6005859375 -0.000539518853503776
10.625 -0.000538462723107725
10.6494140625 -0.000537372960649246
10.673828125 -0.000536249534153017
10.6982421875 -0.000535092412372225
10.72265625 -0.000533901564619073
10.7470703125 -0.000532676960643544
10.771484375 -0.000531418570538681
10.7958984375 -0.000530126364700648
10.8203125 -0.000528800313826099
10.8447265625 -0.000527440388953342
10.869140625 -0.000526046561559488
10.8935546875 -0.000524618803698495
10.91796875 -0.000523157088194919
10.9423828125 -0.000521661388890948
10.966796875 -0.000520131680942018
10.9912109375 -0.00051856794117353
11.015625 -0.000516970148483738
11.0400390625 -0.000515338284313696
11.064453125 -0.000513672333160143
11.0888671875 -0.000511972283152592
11.11328125 -0.000510238126681735
11.1376953125 -0.000508469861082332
11.162109375 -0.00050666748936702
11.1865234375 -0.000504831021014187
11.2109375 -0.000502960472803533
11.2353515625 -0.000501055869699916
11.259765625 -0.000499117245780586
11.2841796875 -0.00049714464520607
11.30859375 -0.000495138123232773
11.3330078125 -0.000493097747256862
11.357421875 -0.000491023597892754
11.3818359375 -0.000488915770087448
11.40625 -0.000486774374250804
11.4306640625 -0.000484599537415261
11.455078125 -0.000482391404410007
11.4794921875 -0.000480150139052703
11.50390625 -0.000477875925349771
11.5283203125 -0.000475568968703391
11.552734375 -0.000473229497121532
11.5771484375 -0.000470857762432283
11.6015625 -0.000468454041488459
11.6259765625 -0.00046601863736737
11.650390625 -0.000463551880567452
11.6748046875 -0.000461054130188543
11.69921875 -0.0004585257751013
11.7236328125 -0.000455967235105441
11.748046875 -0.000453378962077749
11.7724609375 -0.000450761441100689
11.796875 -0.000448115191591473
11.8212890625 -0.000445440768407908
11.845703125 -0.0004427387629621
11.8701171875 -0.000440009804319511
11.89453125 -0.0004372545603015
11.9189453125 -0.000434473738586093
11.943359375 -0.000431668087809555
11.9677734375 -0.000428838398673884
11.9921875 -0.000425985505045295
12.0166015625 -0.0004231102850566
12.041015625 -0.000420213662184316
12.0654296875 -0.000417296606302816
12.08984375 -0.000414360134692292
12.1142578125 -0.000411405312966252
12.138671875 -0.000408433255895227
12.1630859375 -0.000405445128078573
12.1875 -0.000402442144415197
12.2119140625 -0.000399425570309189
12.236328125 -0.000396396721549787
12.2607421875 -0.000393356963772855
12.28515625 -0.000390307711437439
12.3095703125 -0.000387250426213518
12.333984375 -0.000384186614708557
12.3583984375 -0.000381117825437142
12.3828125 -0.000378045644967844
12.4072265625 -0.00037497169319171
12.431640625 -0.000371897617685859
12.4560546875 -0.000368825087160522
12.48046875 -0.000365755784055987
12.5048828125 -0.000362691396359636
12.529296875 -0.000359633608802196
12.5537109375 -0.000356584093636184
12.578125 -0.000353544501252454
12.6025390625 -0.00035051645096084
12.626953125 -0.000347501522291014
12.6513671875 -0.000344501247222935
12.67578125 -0.000341517103753778
12.7001953125 -0.000338550511203393
12.724609375 -0.000335602827645678
12.7490234375 -0.000332675349764667
12.7734375 -0.000329769315353196
12.7978515625 -0.000326885908563085
12.822265625 -0.000324026267840503
12.8466796875 -0.000321191496387721
12.87109375 -0.000318382674752032
12.8955078125 -0.000315600875068129
12.919921875 -0.000312847176296454
12.9443359375 -0.000310122679727256
12.96875 -0.00030742852394329
12.9931640625 -0.000304765898449366
13.017578125 -0.000302136055195074
13.0419921875 -0.000299540317334279
13.06640625 -0.000296980084700653
13.0908203125 -0.000294456835662942
13.115234375 -0.000291972125227158
13.1396484375 -0.000289527579491606
13.1640625 -0.000287124886734583
13.1884765625 -0.00028476578564114
13.212890625 -0.000282452051306802
13.2373046875 -0.00028018547977268
13.26171875 -0.000277967871884612
13.2861328125 -0.000275801017286313
13.310546875 -0.000273686679282203
13.3349609375 -0.000271626581192817
13.359375 -0.00026962239472874
13.3837890625 -0.000267675730668631
13.408203125 -0.000265788132032179
13.4326171875 -0.00026396106968836
13.45703125 -0.000262195940252442
13.4814453125 -0.00026049406593612
13.505859375 -0.000258856695955083
13.5302734375 -0.000257285009028121
13.5546875 -0.000255780116481959
13.5791015625 -0.000254343065495739
13.603515625 -0.000252974842076508
13.6279296875 -0.00025167637341713
13.65234375 -0.000250448529395
13.6767578125 -0.000249292123051611
13.701171875 -0.000248207910015223
13.7255859375 -0.000247196586886616
13.75 -0.000246258788722986
13.7744140625 -0.000245395085783448
13.798828125 -0.000244605979757966
13.8232421875 -0.000243891899730933
13.84765625 -0.000243253198102555
13.8720703125 -0.000242690146720516
13.896484375 -0.000242202933415114
13.9208984375 -0.000241791659106847
13.9453125 -0.000241456335613864
13.9697265625 -0.000241196884249936
13.994140625 -0.000241013135231198
14.0185546875 -0.000240904827894133
14.04296875 -0.000240871611692735
14.0673828125 -0.000240913047888697
14.091796875 -0.000241028611858784
14.1162109375 -0.000241217695906746
14.140625 -0.000241479612486725
14.1650390625 -0.000241813597725074
14.189453125 -0.000242218815161606
14.2138671875 -0.000242694359631026
14.23828125 -0.000243239261221022
14.2626953125 -0.000243852489289983
14.287109375 -0.000244532956499192
14.3115234375 -0.000245279522868909
14.3359375 -0.000246090999873192
14.3603515625 -0.000246966154576726
14.384765625 -0.000247903713839232
14.4091796875 -0.000248902368617113
14.43359375 -0.000249960778358516
14.4580078125 -0.000251077575525069
14.482421875 -0.000252251370210314
14.5068359375 -0.000253480754878646
14.53125 -0.000254764309164122
14.5556640625 -0.000256100604739958
14.580078125 -0.000257488210178359
14.6044921875 -0.000258925695802659
14.62890625 -0.000260411638444712
14.6533203125 -0.000261944626088722
14.677734375 -0.00026352326234472
14.7021484375 -0.000265146170702979
14.7265625 -0.000266811998558169
14.7509765625 -0.000268519420932038
14.775390625 -0.000270267143908357
14.7998046875 -0.000272053907738145
14.82421875 -0.000273878489606425
14.8486328125 -0.00027573970607477
14.873046875 -0.000277636415167295
14.8974609375 -0.00027956751813507
14.921875 -0.000281531960892098
14.9462890625 -0.000283528735132572
14.970703125 -0.000285556879149641
14.9951171875 -0.000287615478358904
15.01953125 -0.000289703665546132
15.0439453125 -0.000291820620845082
15.068359375 -0.000293965571456754
15.0927734375 -0.000296137791129336
15.1171875 -0.000298336599396244
15.1416015625 -0.000300561360584629
15.166015625 -0.00030281148262399
15.1904296875 -0.000305086415633941
15.21484375 -0.000307385650324449
15.2392578125 -0.000309708716216012
15.263671875 -0.000312055179687276
15.2880859375 -0.00031442464186035
15.3125 -0.000316816736346602
15.3369140625 -0.000319231126848468
15.361328125 -0.000321667504652605
15.3857421875 -0.000324125586001776
15.41015625 -0.000326605109380799
15.4345703125 -0.000329105832727259
15.458984375 -0.00033162753056755
15.4833984375 -0.000334169991120046
15.5078125 -0.000336733013349858
15.5322265625 -0.000339316404033407
15.556640625 -0.000341919974803471
15.5810546875 -0.000344543539230559
15.60546875 -0.000347186909934018
15.6298828125 -0.000349849895742231
15.654296875 -0.000352532298929022
15.6787109375 -0.0003552339125176
15.703125 -0.000357954517682064
15.7275390625 -0.000360693881254979
15.751953125 -0.000363451753329477
15.7763671875 -0.000366227864998465
15.80078125 -0.000369021926196971
15.8251953125 -0.000371833623683134
15.849609375 -0.000374662619146437
15.8740234375 -0.000377508547437472
15.8984375 -0.000380371014945254
15.9228515625 -0.000383249598100952
15.947265625 -0.00038614384201763
15.9716796875 -0.000389053259267762
15.99609375 -0.000391977328809508
16.0205078125 -0.000394915495046003
16.044921875 -0.000397867167047887
16.0693359375 -0.000400831717911524
16.09375 -0.000403808484289011
16.1181640625 -0.000406796766066693
16.142578125 -0.000409795826222362
16.1669921875 -0.000412804890830713
16.19140625 -0.000415823149251654
16.2158203125 -0.000418849754474818
16.240234375 -0.000421883823624928
16.2646484375 -0.000424924438643154
16.2890625 -0.000427970647100519
16.3134765625 -0.000431021463172382
16.337890625 -0.000434075868737968
16.3623046875 -0.000437132814616558
16.38671875 -0.00044019122190727
16.4111328125 -0.000443249983438197
16.435546875 -0.000446307965296989
16.4599609375 -0.000449364008436749
16.484375 -0.000452416930342773
16.5087890625 -0.000455465526742843
16.533203125 -0.000458508573337001
16.5576171875 -0.000461544827551648
16.58203125 -0.000464573030283306
16.6064453125 -0.00046759190763229
16.630859375 -0.000470600172593437
16.6552734375 -0.000473596526713478
16.6796875 -0.000476579661685553
16.7041015625 -0.000479548260881553
16.728515625 -0.000482501000796761
16.7529296875 -0.000485436552433672
16.77734375 -0.000488353582578899
16.8017578125 -0.000491250755006669
16.826171875 -0.000494126731593659
16.8505859375 -0.000496980173354506
16.875 -0.000499809741404579
16.8994140625 -0.00050261409786589
16.923828125 -0.000505391906722062
16.9482421875 -0.000508141834643912
16.97265625 -0.000510862551808678
16.9970703125 -0.000513552732715918
17.021484375 -0.000516211057049726
17.0458984375 -0.000518836210578273
17.0703125 -0.000521426886136163
17.0947265625 -0.00052398178470025
17.119140625 -0.000526499616582091
17.1435546875 -0.000528979102749783
17.16796875 -0.000531418976307652
17.1923828125 -0.000533817984124719
17.216796875 -0.000536174888645618
17.2412109375 -0.000538488469861896
17.265625 -0.000540757527463291
17.2900390625 -0.000542980883153101
17.314453125 -0.000545157383125709
17.3388671875 -0.000547285900680923
17.36328125 -0.000549365338965204
17.3876953125 -0.000551394633818654
17.412109375 -0.000553372756686164
17.4365234375 -0.000555298717590731
17.4609375 -0.000557171568105935
17.4853515625 -0.000558990404307685
17.509765625 -0.000560754369690872
17.5341796875 -0.000562462657969727
17.55859375 -0.000564114515768581
17.5830078125 -0.00056570924515064
17.607421875 -0.000567246205957629
17.6318359375 -0.000568724817919913
17.65625 -0.000570144562526189
17.6806640625 -0.000571504984619501
17.705078125 -0.000572805693697906
17.7294921875 -0.000574046364905786
17.75390625 -0.000575226739702783
17.7783203125 -0.000576346626204662
17.802734375 -0.000577405899183674
17.8271484375 -0.000578404499736179
17.8515625 -0.000579342434620025
17.8759765625 -0.000580219775257762
17.900390625 -0.000581036656436734
17.9248046875 -0.000581793274699153
17.94921875 -0.000582489886448124
17.9736328125 -0.000583126805789948
17.998046875 -0.00058370440213411
18.0224609375 -0.000584223097565311
18.046875 -0.000584683364012261
18.0712890625 -0.000585085720256487
18.095703125 -0.00058543072878287
18.1201171875 -0.000585718992500402
18.14453125 -0.00058595115138186
18.1689453125 -0.000586127879017674
18.193359375 -0.000586249879120707
18.2177734375 -0.000586317882016253
18.2421875 -0.000586332641114992
18.2666015625 -0.00058629492942043
18.291015625 -0.000586205536057049
18.3154296875 -0.000586065262871186
18.33984375 -0.000585874921093359
18.3642578125 -0.000585635328098106
18.388671875 -0.000585347304262745
18.4130859375 -0.000585011669944471
18.4375 -0.000584629242578016
18.4619140625 -0.000584200833917431
18.486328125 -0.00058372724741688
18.5107421875 -0.000583209275765316
18.53515625 -0.000582647698574493
18.5595703125 -0.000582043280230976
18.583984375 -0.0005813967679105
18.6083984375 -0.000580708889763167
18.6328125 -0.00057998035325922
18.6572265625 -0.00057921184371007
18.681640625 -0.00057840402295726
18.7060546875 -0.000577557528220205
18.73046875 -0.000576672971121222
18.7548828125 -0.000575750936865472
18.779296875 -0.000574791983583037
18.8037109375 -0.000573796641829221
18.828125 -0.000572765414236624
18.8525390625 -0.000571698775314127
18.876953125 -0.00057059717139444
18.9013671875 -0.00056946102071847
18.92578125 -0.000568290713651821
18.9501953125 -0.00056708661303262
18.974609375 -0.000565849054645105
18.9990234375 -0.000564578347805791
19.0234375 -0.000563274776065039
19.0478515625 -0.000561938598011561
19.072265625 -0.000560570048181813
19.0966796875 -0.00055916933806426
19.12109375 -0.00055773665718179
19.1455078125 -0.000556272174276011
19.169921875 -0.000554776038544535
19.1943359375 -0.000553248380976618
19.21875 -0.00055168931573183
19.2431640625 -0.000550098941592658
19.267578125 -0.000548477343469137
19.2919921875 -0.000546824593951609
19.31640625 -0.000545140754910239
19.3408203125 -0.000543425879134207
19.365234375 -0.000541680012007239
19.3896484375 -0.000539903193209929
19.4140625 -0.000538095458448998
19.4384765625 -0.000536256841209904
19.462890625 -0.000534387374520457
19.4873046875 -0.000532487092726449
19.51171875 -0.000530556033283089
19.5361328125 -0.000528594238533172
19.560546875 -0.000526601757496808
19.5849609375 -0.000524578647649244
19.609375 -0.000522524976686969
19.6337890625 -0.000520440824284412
19.658203125 -0.000518326283826938
19.6826171875 -0.000516181464126775
19.70703125 -0.000514006491111259
19.7314453125 -0.000511801509475028
19.755859375 -0.000509566684312133
19.7802734375 -0.000507302202700335
19.8046875 -0.000505008275252416
19.8291015625 -0.000502685137616277
19.853515625 -0.000500333051939356
19.8779296875 -0.000497952308274195
19.90234375 -0.000495543225929566
19.9267578125 -0.000493106154771614
19.951171875 -0.000490641476455037
19.9755859375 -0.000488149605594942
20 -0.000485630990861601
};
\end{axis}

\end{tikzpicture}